\documentclass[10pt,twocolumn,twoside,journal]{IEEEtran}

\usepackage{subfigure}
\usepackage{multirow}
\usepackage{psfig}
\usepackage{graphicx}
\usepackage{amsmath}
\usepackage{color}
\usepackage{epsfig}
\usepackage{amsfonts}
\usepackage{amssymb}
\usepackage{amsthm}
\usepackage{algorithm2e}
\usepackage{algorithmic}
\usepackage[usenames,dvipsnames]{pstricks}
\usepackage{epstopdf}
\usepackage{arydshln}

\allowdisplaybreaks
\newcommand{\Define}{\triangleq}

\newcommand{\argmax}{\mathop{\text{argmax}}}
\newcommand{\argmin}{\mathop{\text{argmin}}}

\newcommand{\vh}{{\bf h}}

\newcommand{\vx}{{\bf x}}
\newcommand{\vy}{{\bf y}}
\newcommand{\vs}{{\bf s}}
\newcommand{\vn}{{\bf n}}

\newcommand{\vv}{{\bf v}}

\newcommand{\sa}{{\mathbb A}}

\begin{document}

\title{{\huge On Media-based Modulation using RF Mirrors}}
\author{Y. Naresh and A. Chockalingam 

Department of ECE, Indian Institute of Science, Bangalore 560012 
}
\maketitle

\thispagestyle{empty}
\begin{abstract}
Media-based modulation (MBM) is a recently proposed modulation scheme 
which uses radio frequency (RF) mirrors at the transmit antenna(s) in 
order to create different channel fade realizations based on their ON/OFF 
status. These complex fade realizations constitute the modulation alphabet. 
MBM has the advantage of increased spectral efficiency and performance.  
In this paper, we investigate the performance of some physical layer
techniques when applied to MBM. Particularly, we study the performance of 
$i)$ MBM with generalized spatial modulation (GSM), $ii)$ MBM with mirror 
activation pattern (MAP) selection based on an Euclidean distance (ED) 
based metric, and $iii)$ MBM with feedback based phase compensation and 
constellation rotation. Our results show that, for the same spectral 
efficiency, GSM-MBM can achieve better performance compared to MIMO-MBM. 
Also, it is found that MBM with ED-based MAP selection results in improved 
bit error performance, and that phase compensation and MBM constellation 
rotation increases the ED between the MBM constellation points and 
improves the performance significantly. We also analyze the diversity 
orders achieved by the ED-based MAP selection scheme and the phase 
compensation and constellation rotation (PC-CR) scheme. The diversity 
orders predicted by the analysis are validated through simulations.
\end{abstract}

{\em {\bfseries keywords:}}
{\em {\footnotesize
Media-based modulation (MBM), RF mirrors, mirror activation pattern (MAP),
MIMO-MBM, GSM-MBM, MAP selection, phase compensation, constellation
rotation.}}

\pagestyle{empty}
\section{Introduction}
Traditionally, symbols chosen from complex modulation alphabets like 
quadrature amplitude modulation (QAM) and phase shift keying (PSK) 
are used to convey information bits, and complex fades 
introduced by the channel are viewed as detrimental effects that 
cause amplitude and phase distortion to the transmitted symbols.
An alternate and interesting approach is to consider the complex 
channel fade coefficients themselves to constitute a modulation 
alphabet. One simple and known example of this approach is space 
shift keying (SSK) \cite{ssk1},\cite{ssk2}, which can be briefly 
explained as follows. 

\subsubsection{SSK}
Suppose there are two transmit antennas and one receive antenna. 
Assume rich scattering. Let $h_1 \sim {\cal CN}(0,1)$ and 
$h_2 \sim {\cal CN}(0,1)$ denote the complex channel fade 
coefficients from transmit antennas 1 and 2, respectively, to the 
receive antenna.  Now, assuming that a tone of unit amplitude is 
transmitted by any one of the transmit antennas in a given channel
use, ${\mathbb H}\Define \{h_1,h_2\}$ can be viewed as the underlying 
modulation alphabet, i.e., $h_1$ and $h_2$ are the {\em random} 
constellation points. The alphabet ${\mathbb H}$ therefore can convey 
$\log_2|{\mathbb H}|=\log_22=1$ information bit. To realize this,
if the information bit is 0, antenna 1 transmits the tone and 
antenna 2 remains silent, and if the information bit is 1, 
antenna 1 remains silent and antenna 2 transmits the tone. For
this, it is enough to have only one transmit radio frequency (RF) 
chain, whose output is switched to either antenna 1 or antenna 2
depending on the information bit being 0 or 1, respectively. The 
modulation alphabet (i.e., ${\mathbb H}$) needs to be known at 
the receiver for detection, which can be obtained through pilot 
transmission and channel estimation. Note, however, that the 
transmitter need not know the alphabet ${\mathbb H}$. The 
information bit is detected using the estimated ${\mathbb H}$ 
at the receiver. 

Similar to the binary SSK scheme with $|{\mathbb H}|=2$ described 
above, higher-order SSK with $|{\mathbb H}|=n_t$ and 
${\mathbb H}=\{h_1,h_2,\cdots,h_{n_t}\}$ can be realized using 
$n_t=2^m$ transmit antennas, and sending the tone on an antenna 
chosen based on $m$ information bits. Therefore, using $n_t=2^m$ 
transmit antennas, SSK achieves a throughput of $m=\log_2n_t$ bits 
per channel use (bpcu). 

If the  receiver has $n_r$ receive antennas, then the alphabet 
${\mathbb H}$ will consist of vector constellation points, i.e., 
${\mathbb H}=\{{\mathbf h}_1,{\mathbf h}_2,\cdots,{\mathbf h}_{n_t}\}$,
where ${\mathbf h}_j=[h_{1,j} \ h_{2,j} \ \cdots \ h_{n_r,j}]^T$, and 
$h_{i,j} \sim \mathcal{CN}(0,1)$ is the channel fade coefficient from 
$j$th transmit antenna to $i$th receive antenna (see Fig. \ref{sskfig}). 
Because of the increased dimensionality of the constellation points for 
increasing $n_r$, the performance of SSK improves significantly 
with increasing $n_r$.  SSK has the advantages of $i)$ requiring  
only one transmit RF chain and a $1\times n_t$ RF switch for any $n_t$, 
and $ii)$ yielding attractive performance at higher spectral 
efficiencies. A key drawback, however, is that SSK needs exponential 
increase in number of transmit antennas to increase the spectral 
efficiency. For example, to achieve $m=8$ bpcu, SSK requires 
$n_t=2^8=256$ transmit antennas. This drawback of the need to have a 
large number of transmit antennas to increase the spectral efficiency 
is significantly alleviated in the recently proposed media-based 
modulation (MBM) scheme \cite{mbm1}-\cite{mbm3}, realized through 
the use of RF mirrors which are turned ON/OFF\footnote{An RF mirror 
in ON status implies that the mirror allows the incident wave to 
pass through it transparently, and an OFF status implies that the mirror 
reflects back the incident wave.} on a channel use-by-channel use basis 
depending on information bits to `modulate' the fade coefficients of the 
channel \cite{mbm3}. The basic MBM can be briefly explained as follows.

\subsubsection{MBM}
The basic version of MBM, like SSK, transmits a tone and uses the 
complex channel fade realizations themselves as the modulation alphabet 
\cite{mbm1}-\cite{mbm2a}. While multiple transmit antennas are 
needed to create the complex fade symbols of the alphabet in SSK, MBM 
creates the complex fade symbols of the alphabet even with a single 
transmit antenna through the use of multiple RF mirrors \cite{mbm3}.
This is achieved by placing a number of RF mirrors near the transmit 
antenna which transmits a tone.  
Placing RF mirrors near a transmit antenna is equivalent to placing 
scatterers in the propagation environment close to the transmitter.
The radiation characteristics of each of these scatterers (i.e., RF 
mirrors) can be changed by an ON/OFF control signal applied to it; it
reflects back the incident wave originating from the transmit antenna
or passes the wave depending on whether it is OFF or ON, respectively. 
We call the 
ON/OFF status of the mirrors as the `mirror activation pattern'. The 
positions of the ON mirrors and OFF mirrors change from one mirror 
activation pattern (MAP) to the other, i.e., the propagation environment 
close to the transmitter changes from one MAP to the other MAP. Note that 
in a rich scattering environment, since a small perturbation in the 
propagation environment will be augmented by many random reflections 
resulting in an independent channel. The RF mirrors create such 
perturbations by acting as controlled scatterers, which, in turn,  
create independent fade realizations for different MAPs.

\begin{figure}
\centering
\subfigure[SSK]{
\includegraphics[height=1.25in]{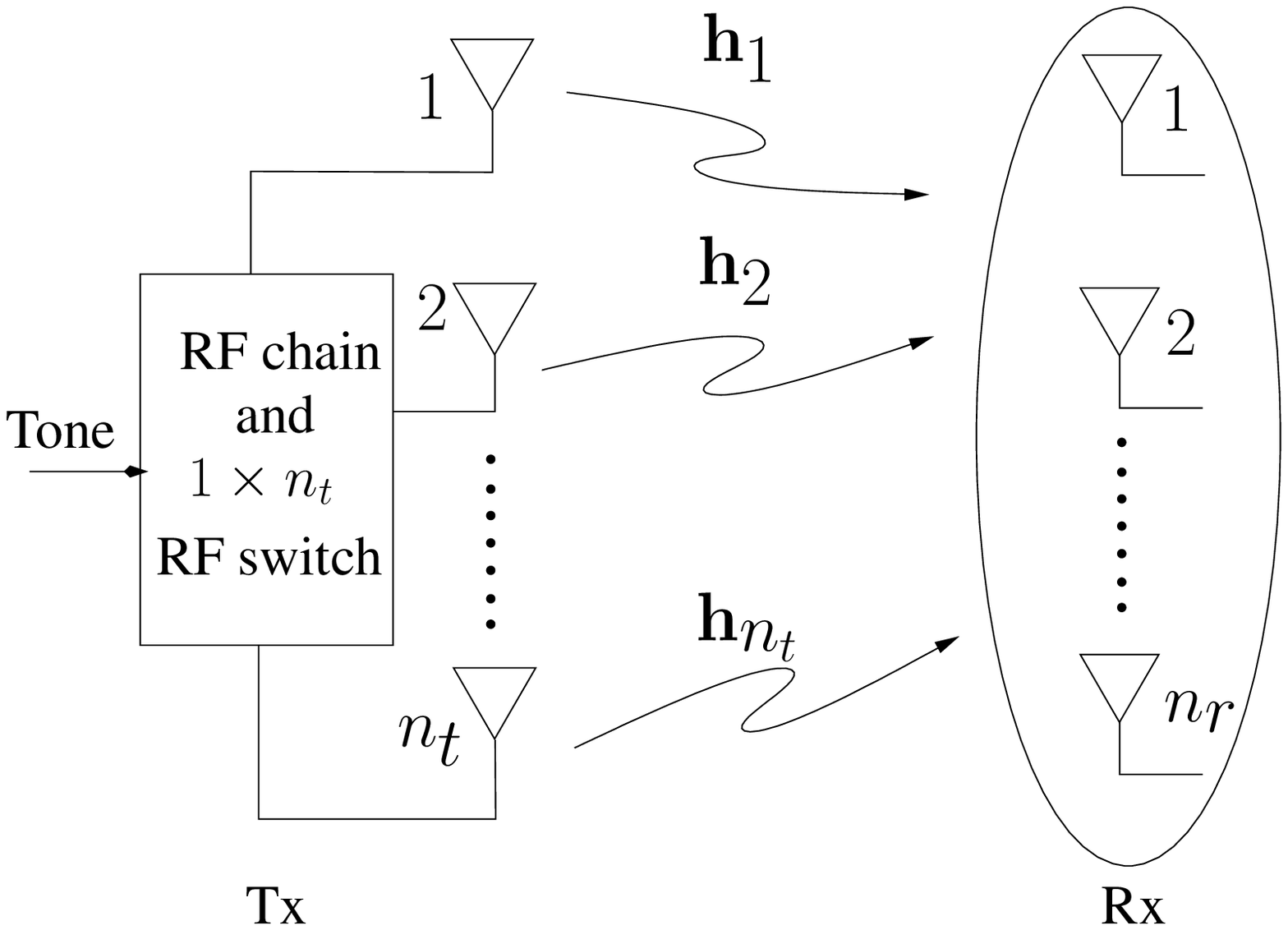}
\label{sskfig}
}
\hspace{3mm}
\subfigure[MBM]{
\includegraphics[height=1.25in]{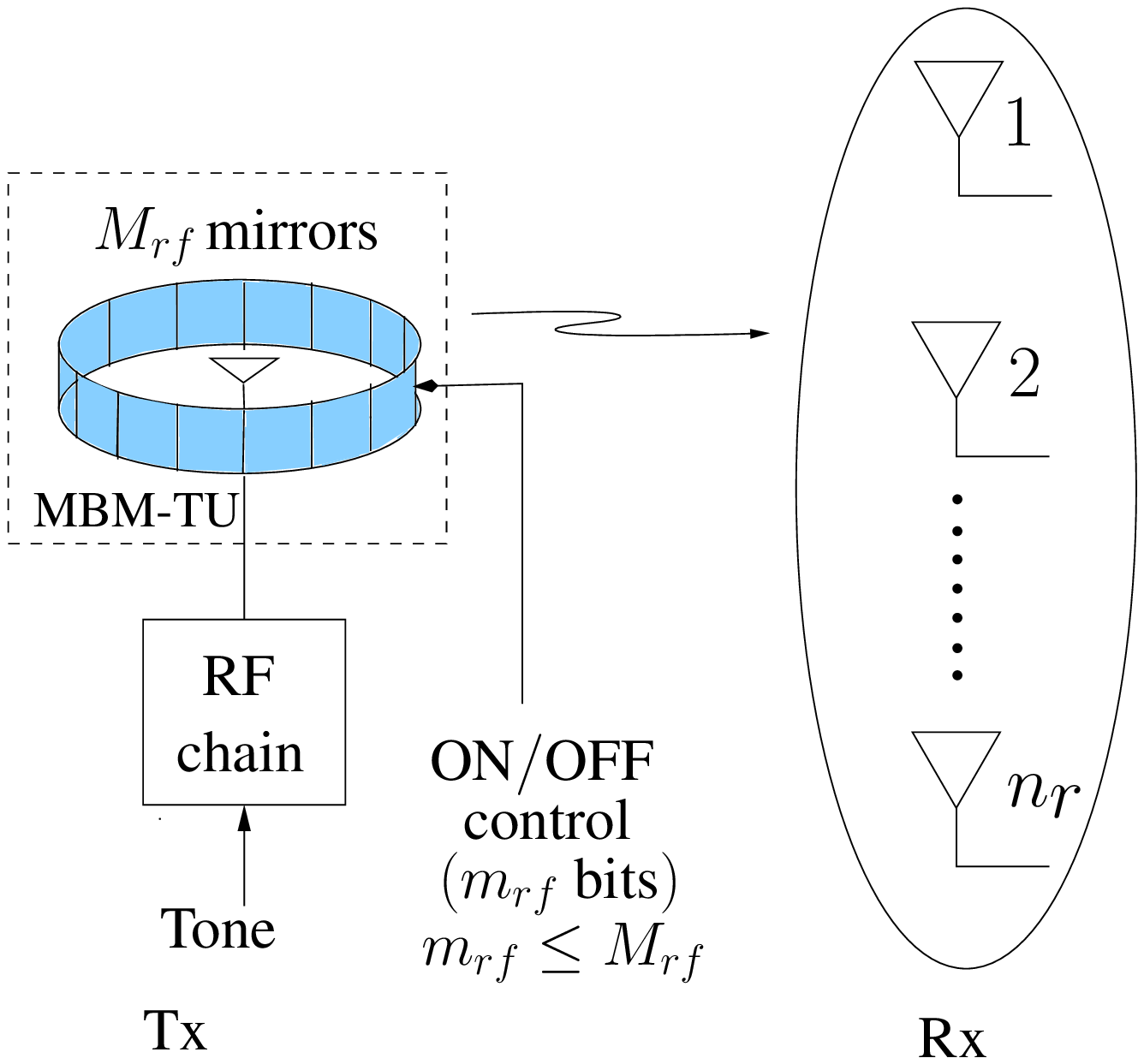}
\label{mbmfig}
}
\caption{(a) SSK with constellation 
${\mathbb H}_{\mbox{ssk}}=\{{\mathbf h}_1, {\mathbf h}_2, \cdots, {\mathbf h}_{n_t}\}$. 
(b) MBM with constellation 
${\mathbb H}_{\mbox{mbm}}=\{{\mathbf h}_1, {\mathbf h}_2, \cdots, {\mathbf h}_{2^{m_{rf}}}\}$. }
\vspace{-4mm}
\label{ssk_mbm}
\end{figure}

Consider a single transmit antenna. Let $M_{rf}$ denote the total number 
of RF mirrors placed near the antenna. We call the unit comprising of a 
transmit antenna and the set of $M_{rf}$ RF mirrors associated with it 
as the `MBM transmit unit (MBM-TU)' (see Fig. \ref{mbmfig}). Out of the 
$M_{rf}$ available RF mirrors, let $m_{rf}$, where $1\leq m_{rf}\leq M_{rf}$, 
denote the number of RF mirrors actually used. Each of these $m_{rf}$ 
mirrors is turned ON or OFF in a given channel use based on one information 
bit. A realization of the ON/OFF status of all the $m_{rf}$ mirrors, 
determined by $m_{rf}$ information bits, is called a mirror activation 
pattern (MAP). Therefore, $2^{m_{rf}}$ MAPs are possible. Each of these 
patterns results in a different realization of the channel fade, resulting 
in an MBM alphabet of size $|{\mathbb H}|=2^{m_{rf}}$. Therefore, MBM can 
convey $m_{rf}$ information bits in one channel use, where $m_{rf}$ is the 
number of RF mirrors used. That is, the spectral efficiency 
in MBM scales linearly  with the number of RF mirrors used. 
Note that the spectral efficiency in SSK scales logarithmically in the 
number of transmit antennas $n_t$, given by $\lfloor \log_2n_t\rfloor$ bpcu. 
In generalized SSK (GSSK) \cite{gssk}, where $n_{rf}$ out of $n_t$ antennas 
are activated using $n_{rf}$ RF chains, the spectral efficiency increases 
more than logarithmically and less than linearly in $n_t$, given by 
$\lfloor \log_2{n_t\choose n_{rf}}\rfloor$ bpcu. As in SSK, in MBM also 
the alphabet ${\mathbb H}$ needs to be known at the receiver and not at 
the transmitter. 

MBM has been shown to achieve significant performance gains compared to 
conventional modulation schemes \cite{mbm1}-\cite{mbm2a}. Even with a 
single transmit antenna and $n_r$ receive antennas, MBM has been shown 
to achieve significant energy savings compared to a conventional 
$n_r\times n_t$ MIMO system with $n_r=n_t$ \cite{mbm1}. It has also been 
shown that MBM with 1 transmit and $n_r$ receive antennas over a static 
multipath channel asymptotically achieves the capacity of $n_r$ parallel 
AWGN channels \cite{mbm2}. Implementation of an MBM-TU consisting of 
$M_{rf}=14$ RF mirrors placed in a compact cylindrical structure with 
a dipole transmit antenna element placed at the center of the cylindrical 
structure has been reported in \cite{mbm3}. 

We note that the MBM alphabet ${\mathbb H}$ has to be estimated a priori 
at the receiver. This is achieved through pilot transmissions. Since the 
number of complex fade symbols to be estimated is $n_{tu}2^{m_{rf}}$, the 
number of pilot channel uses needed grows exponentially in $m_{rf}$. 
Channel sounding to learn the alphabet a priori, therefore, is a key 
issue in MBM.

\subsubsection{SM, GSM, MIMO}
The need to exponentially increase the number of transmit antennas to 
increase the spectral efficiency in SSK can be alleviated by allowing 
the transmission of an $M$-ary QAM/PSK symbol on the chosen antenna 
instead of a tone. This allows an additional $\log_2M$ bits to be 
conveyed per channel use by the QAM/PSK symbol. This scheme is called 
the spatial modulation (SM) scheme, which achieves a spectral efficiency 
of $\log_2n_t+\log_2M$ bpcu \cite{sm1},\cite{lajos},\cite{lmimo}.
A further generalization is to use more than one transmit RF chain
(say, $n_{rf}$ transmit RF chains, $1\leq n_{rf} \leq n_t$), and 
transmit $n_{rf}$ QAM/PSK symbols through these RF chains. This scheme 
is called as the generalized spatial modulation (GSM) scheme, whose
spectral efficiency is given by 
$\lfloor \log_2{n_t\choose n_{rf}}\rfloor+n_{rf}\log_2M$ bpcu
\cite{gsm0},\cite{gsm1},\cite{gsm2}. For $n_{rf}=n_t$, the GSM scheme
specializes to the well known MIMO (spatial multiplexing) scheme, whose
spectral efficiency is given by $n_t\log_2M$ bpcu. SM systems in 
frequency-selective channels have been studied in \cite{sc1}-\cite{sc3}. 
The design of SM systems for frequency-selective channels has been summarized 
in \cite{yang_sc}, which also discusses SM variants like single-carrier 
SM (SC-SM), SC-GSM, orthogonal frequency division multiplexing SM (OFDM-SM), 
space and time-dispersion modulation (STdM), and space-frequency shift 
keying (SFSK).

\subsubsection{SM-MBM, GSM-MBM, MIMO-MBM}
Multiple MBM-TUs (i.e., multiple transmit antennas, each having its 
own set of RF mirrors) can be used to increase the spectral efficiency
in MBM. Also, like QAM/PSK symbols (as in SM) and additional transmit 
RF chains (as in GSM, MIMO) could be used to increase the 
spectral efficiency beyond that can be achieved using SSK, 
one could use a similar approach to increase the spectral 
efficiency 
beyond that can be  achieved using the basic MBM. Spatial modulation when 
used with MBM is referred to as the SM-MBM scheme, whose spectral efficiency 
is given by $m_{rf} + \lfloor \log_2{n_{tu}}\rfloor + \log_2M$ 
bpcu, where $n_{tu}$ is the number of MBM-TUs. Similarly, GSM and MIMO 
(spatial multiplexing) when used with MBM are called GSM-MBM and MIMO-MBM, 
respectively. The spectral efficiency of GSM-MBM is given by 
$n_{rf}m_{rf}+\lfloor \log_2{n_{tu}\choose n_{rf}}\rfloor+n_{rf}\log_2M$
bpcu. The spectral efficiency of MIMO-MBM is given by
$n_{tu}m_{rf} + n_{tu}\log_2M$ bpcu. 
Note that GSM-MBM  specializes to SM-MBM when $n_{tu}>1$ 
and $n_{rf}=1$. In each channel use, GSM-MBM has two levels of indexing, 
namely, 1) MBM-TU indexing: $n_{rf}$ out of $n_{tu}$ MBM-TUs are selected 
using $\lfloor\log_2{n_{tu}\choose n_{rf}}\rfloor$ bits, and 2) RF mirror 
indexing: a MAP (ON/OFF status of the mirrors) is selected in each of the 
selected MBM-TUs using $m_{rf}$ bits. The performance of MIMO-MBM
has been studied in \cite{mbm3}, where it has been shown that MIMO-MBM
can achieve better performance compared to conventional MIMO. GSM-MBM
performance has not been studied in \cite{mbm3}. One of our contributions 
in this paper is to present the performance of GSM-MBM and compare it 
with that of MIMO-MBM.

\subsubsection{MBM viewed as an instance of index modulation}
MBM can be viewed as an instance of index modulation, where
information bits are conveyed through the indices of certain transmit 
entities that get involved in the transmission. Indexing transmit 
antennas in multi-antenna systems (e.g., SSK, SM, GSM 
\cite{ssk1},\cite{ssk2},\cite{sm1}-\cite{gsm2}), indexing subcarriers 
in multi-carrier systems \cite{scim1}-\cite{scim2a}, indexing 
both transmit antennas and subcarriers \cite{gsfm}, and indexing
precoders \cite{pim1} are examples of such instances. In this context,
MBM also can be viewed as an index modulation scheme, where RF mirrors 
act as the transmit entities that are indexed to convey information bits.
In SM-MBM and GSM-MBM, indexing is done both on the MBM-TUs as well as
the RF mirrors in each of the chosen MBM-TUs for transmission. 

\subsubsection{Parasitic elements and reconfigurable antennas}
\label{subsec_parasite}
The idea of using parasitic elements external to the antenna (which may 
include capacitors, varactors or switched capacitors that can adjust the 
resonance frequency) for the purpose of creating multiple antenna patterns 
is widely known in the literature \cite{schaer}-\cite{reconfig_2}. By 
adjusting the resonance frequency of the different parasitic elements, 
different channel states for the signal radiating from the antenna can 
be realized. The use of parasitic elements 
for beamforming purposes (i.e., to focus the radiated RF energy in specific 
directions) has been widely studied \cite{schaer}. Other applications of 
parasitic elements reported in the literature include selection/switched 
diversity \cite{vaug99}, base station tracking using switched parasitic 
antenna array \cite{base_track}, direction of arrival estimation with a 
single-port parasitic array antenna \cite{doa_est}, and reconfigurable 
antennas \cite{reconfig_1},\cite{reconfig_2}. These applications do not 
index antenna patterns to convey information bits, i.e., they do not use 
parasitic elements for index modulation purposes. Termed as `aerial 
modulation,' the idea of indexing orthogonal antenna patterns (realized 
using a single antenna surrounded by parasitic elements) to convey 
information bit(s) in addition to bits conveyed through $M$-PSK symbol 
has been studied in \cite{kalis1},\cite{kalis2}. The idea of conveying 
information bits through antenna pattern indexing has also been 
highlighted in \cite{R_bains}. MBM in \cite{mbm1}-\cite{mbm2a} also, in 
a similar way, exploits the idea of indexing a multiplicity of channel 
gain profiles realized by placing RF mirrors around the antenna and 
allowing information bits to control the transparent/opaque status of 
these mirrors. Extension of MBM to multiple transmit antennas is studied 
in \cite{mbm3}. In addition, \cite{mbm1}-\cite{mbm3} also highlight 
the advantages of such systems like additive properties of information 
over multiple receive antennas.

\subsubsection{Contributions in this paper}
In this paper, we investigate the performance of some physical layer 
techniques when applied to MBM \cite{ita2016}. These techniques include 
generalized spatial modulation, MAP selection (analogous to antenna 
selection in MIMO systems), and phase compensation and constellation 
rotation (PC-CR). Our contributions in this paper can be summarized as 
follows.  
\begin{itemize}
\item 	We study the performance of MBM with generalized spatial modulation 
	(referred to as GSM-MBM), and compare its performance 
        with that of MIMO-MBM. Our results show that, for the same spectral 
        efficiency, GSM-MBM can achieve better performance compared to 
	MIMO-MBM. A union bound based upper bound on the 
	average bit error probability (BEP) of GSM-MBM is shown to be tight 
	for moderate-to-high SNRs. 
\item 	We investigate a MAP selection scheme based on an Euclidean 
	distance (ED) based metric, and compare its bit error performance 
	with that of a mutual information (MI) based MAP selection scheme. 
        The ED-based MAP selection scheme is found to perform better
	than the MI-based MAP selection scheme by several dBs. The 
	diversity order achieved by the ED-based MAP selection scheme 
	is shown to be $n_r(2^{M_{rf}}-2^{m_{rf}}+1)$, which is 
        validated through simulations as well.
\item 	We investigate a scheme with feedback based phase compensation 
        and MBM constellation rotation, which increases the ED between 
	the constellation points and improves the bit error performance 
	significantly. The diversity order achieved by this 
	scheme is shown to be $n_r(n_{tu}+1)$, which is also validated 
	through simulations.
\end{itemize}
The rest of this paper is organized as follows. Section 
\ref{sec2} presents the GSM-MBM scheme. Section \ref{sec3} presents 
the ED-based MAP selection scheme and its diversity analysis. Section 
\ref{sec4} presents the feedback based PC-CR scheme and its diversity 
analysis. Section \ref{sec5} presents the results and discussions. 
Conclusions are presented in \ref{sec6}.

\section{GSM-MBM scheme}
\label{sec2}
\begin{figure}
\centering
\includegraphics[height=2.75in,width=3.25in]{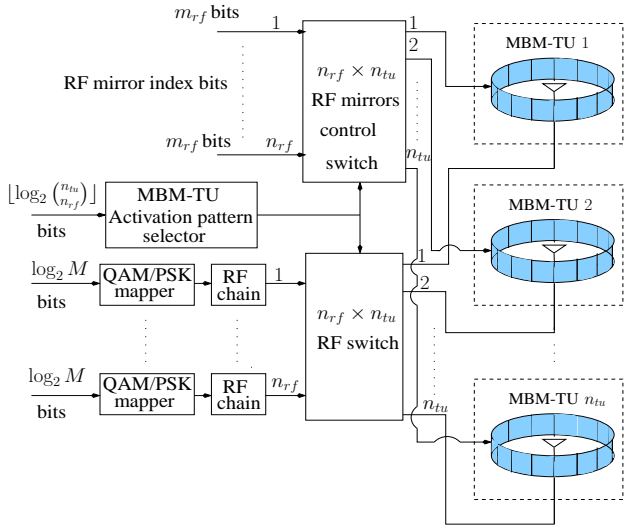}
\caption{GSM-MBM transmitter.}
\label{gsm_mbm_fig}
\vspace{-4mm}
\end{figure}

In this section, we introduce the GSM-MBM scheme and analyze its bit 
error performance. The GSM-MBM transmitter is shown in Fig. 
\ref{gsm_mbm_fig}. It consists of $n_{tu}$ MBM-TUs, $n_{rf}$ transmit 
RF chains, $1\leq n_{rf} \leq n_{tu}$, and an $n_{rf}\times n_{tu}$ 
RF switch. In each MBM-TU, $m_{rf}$ RF mirrors are used. In GSM-MBM,
information bits are conveyed using MBM-TU indexing, RF mirror 
indexing, and QAM/PSK symbols, as follows. In each channel use, 
$i)$ $n_{rf}$ out of $n_{tu}$ MBM-TUs are selected using 
$\lfloor\log_2{n_{tu}\choose n_{rf}}\rfloor$ bits, $ii)$ on the
selected $n_{rf}$ MBM-TUs, $n_{rf}$ $M$-ary QAM/PSK symbols (formed 
using $n_{rf}\log_2M$ bits) are transmitted, and $iii)$ the ON/OFF 
status of the $m_{rf}$ mirrors in each of the selected MBM-TU is 
controlled by $m_{rf}$ bits (so that all the $n_{rf}m_{rf}$ mirrors 
in the selected $n_{rf}$ MBM-TUs are controlled by $n_{rf}m_{rf}$
bits). Therefore, the achieved rate in bpcu is given by
\begin{equation}
\eta = \underbrace{\Big\lfloor\log_2{n_{tu}\choose n_{rf}}\Big\rfloor}_{\mbox{{\scriptsize MBM-TU index bits}}} + \hspace{-1mm} \underbrace{n_{rf}m_{rf}}_{\mbox{{\scriptsize mirror index bits}}} \hspace{-1mm} + \hspace{-1mm} \underbrace{n_{rf}\log_2M}_{\mbox{\scriptsize QAM/PSK symbol bits}} \hspace{-1mm} \mbox{bpcu}.
\end{equation}
GSM-MBM specializes to SIMO-MBM when $n_{tu}=n_{rf}=1$, to SM-MBM when 
$n_{tu}>1$ and $n_{rf}=1$, and to MIMO-MBM when $n_{tu}>1$ and 
$n_{rf}=n_{tu}$.

\subsection{GSM signal set}
Let ${\mathbb A}$ denote the $M$-ary QAM/PSK alphabet used. In a given
channel use, each of the $n_{tu}$ MBM-TUs is made ON (and a symbol from 
${\mathbb A}$ is sent) or OFF (which is equivalent to sending 0) 
based on information bits. Let us call a realization of the ON/OFF 
status of the $n_{tu}$ MBM-TUs as a `MBM-TU activation pattern'. A total 
of ${n_{tu}\choose n_{rf}}$ MBM-TU activation patterns are possible. 
Out of them, only $2^{\lfloor\log_2{n_{tu}\choose n_{rf}}\rfloor}$ are 
needed for signaling. Let ${\mathbb S}_t$ denote the set of these 
$2^{\lfloor\log_2{n_{tu}\choose n_{rf}}\rfloor}$ MBM-TU activation 
patterns chosen from the set of all possible MBM-TU activation 
patterns. A mapping is done between combinations of 
${\lfloor\log_2{n_{tu}\choose n_{rf}}\rfloor}$ bits to MBM-TU 
activation patterns in ${\mathbb S}_t$. The GSM signal set, denoted 
by ${\mathbb S}_{\mbox{{\scriptsize gsm}}}$, is the set of 
$n_{tu}\times 1$-sized GSM signal vectors that can be transmitted,
which is given by 
\begin{eqnarray}
{\mathbb S}_{\mbox{{\scriptsize gsm}}} = \big \{ \vs :  s_j \in \sa\cup\{0\}, \, \lVert\vs\rVert_0=n_{rf}, \, {\cal I}(\vs)\in {\mathbb S}_t \big \},
\end{eqnarray}
where $\vs$ is the $n_{tu}\times 1$ transmit vector, $s_j$ is the $j$th
entry of $\vs$, $j=1,2,\cdots,n_{tu}$, $\lVert\vs\rVert_0$ is the 
$l_0$-norm of the vector $\vs$, and ${\cal I}(\vs)$ is a function 
that gives the MBM-TU activation pattern for $\vs$. For example, 
when ${\mathbb A}$ is BPSK, $n_{tu}=4$ and $n_{rf}=3$, 
$\vs=[+1 \ \ 0  \ -1 \ -1]^T$ is a valid GSM signal vector, and 
$\mathcal{I}(\vs)$ in this case is given by
${\cal I}(\vs=[+1 \ \ 0  \ -1 \ -1]^T)=[1 \ 0 \ 1 \ 1]^T$. 

\subsection{GSM-MBM received signal}
In GSM-MBM, in addition to the bits conveyed by the GSM signal vector, 
the channel fade symbols created by the RF mirrors in the active 
MBM-TUs also convey additional bits. Let $n_r$ denote the number of 
receive antennas. Since $m_{rf}$ mirrors are used in each MBM-TU, the 
number of mirror activation patterns (MAPs) on each MBM-TU is given by 
$N_m=2^{m_{rf}}$. Let ${\mathbb S}_m$ denote the set of all $N_m$ MAPs 
per MBM-TU. Let ${\mathbb H}^j$ denote the MBM alphabet of size $N_m$, 
consisting of $n_r\times 1$-sized vector constellation points formed 
using the channel fade coefficients corresponding to the $N_m$ MAPs 
of $j$th MBM-TU to the receive antennas. Let 
${\mathbf h}_{k}^{j}=[h_{1,k}^j \ h_{2,k}^j \ \cdots \ h_{n_r,k}^j]^T$
denote the $n_r\times 1$-sized channel coefficient vector at the 
receiver for the $k$th MAP of the $j$th MBM-TU, where 
$h_{i,k}^j$ is the fade coefficient 
corresponding to the $k$th MAP of $j$th MBM-TU to the $i$th receive 
antenna, $i=1,2,\cdots,n_{r}$, $j=1,2,\cdots,n_{tu}$, and 
$k=1,2,\cdots,N_m$.
The $h_{i,k}^j$s are i.i.d. and distributed as $\mathcal{CN}(0,1)$.
We then have ${\mathbb H}^j=
\{{\mathbf h}_{1}^{j},{\mathbf h}_{2}^{j},\cdots,{\mathbf h}_{N_m}^{j}\}$. 
Let $s_j\in {\mathbb A}$ denote the $M$-ary QAM/PSK symbol transmitted
on the $j$th MBM-TU. The received signal vector 
${\mathbf y}=[y_1 \ y_2 \ \cdots \ y_{n_r}]^T$ in a given channel use is 
then given by
\begin{equation}
{\mathbf y}=\sum_{j=1}^{n_{tu}} s_j {\mathbf h}_{l_j}^{j}+{\mathbf n}, 
\label{sys1}
\end{equation}
where 
$s_j\in {\mathbb A}\cup \{0\}$, $l_j\in\{1,\cdots,N_m\}$ is the index of 
the MAP chosen on the $j$th MBM-TU, 
${\mathbf h}_{l_j}^{j}\in {\mathbb H}^j$, and 
${\mathbf n}=[n_1 \ n_2 \ \cdots \ n_{n_r}]^T$ is the additive 
noise vector, whose elements are i.i.d. and distributed as 
$\mathcal{CN}(0,\sigma^2)$. Let 
${\bf H}^j=[\vh_1^j \ \vh_2^j \ \cdots \ \vh_{N_m}^j]$ denote the
$n_r\times N_m$ channel matrix corresponding to the $j$th MBM-TU.
The received vector $\vy$ in (\ref{sys1}) can be written as
\begin{equation}
\vy = \sum_{j=1}^{n_{tu}} s_j{\mathbf H}^j{\mathbf e}_{l_j} + \vn,
\end{equation}
where ${\mathbf e}_{l_j}$ is an $N_m\times 1$ vector whose $l_j$th
coordinate is 1 and all other coordinates are zeros.
Now, defining 
${\mathbf H}=[{\mathbf H}^1 \ {\mathbf H}^2 \ \cdots \ {\mathbf H}^{n_{tu}}]$
as the overall $n_r\times N_mn_{tu}$ channel matrix, we can write 
${\bf y}$ as
\begin{equation}
\vy = \mathbf{Hx}+\vn,
\label{syseq1}
\end{equation}
where $\mathbf x$ belongs to the GSM-MBM signal set
${\mathbb S}_{\mbox{{\scriptsize gsm-mbm}}}$, which is given by
\begin{eqnarray}
\hspace{-5mm}
{\mathbb S}_{\mbox{{\scriptsize gsm-mbm}}} 
\hspace{-2mm}&=&\hspace{-2mm} \left\{ 
{\mathbf x}= [\mathbf{x}_1^T \ \mathbf{x}_2^T \ \cdots \ \mathbf{x}_{n_{tu}}^T]^T : \mathbf{x}_j=s_j\mathbf{e}_{l_j}, 
\right. \nonumber \\
& & \left. \hspace{-1.4cm} l_j \in \{1,\cdots,N_m\}; \ \ 
\vs=[s_1 \ s_2 \ \cdots \ s_{n_{tu}}]^T \in {\mathbb S}_{\mbox{{\scriptsize gsm}}} \right\}.
\label{gsm_set}
\end{eqnarray}
The maximum likelihood (ML) decision rule is given by
\begin{equation}
\hat{\vx}=\argmin_{{\vx} \in {\mathbb S}_{\mbox{{\scriptsize gsm-mbm}}}} 
\|\vy - \mathbf{Hx}\|^2.
\label{syseq2}
\end{equation} 
The bits corresponding to $\hat{\vx}$ are demapped as follows: 
$i)$ the MBM-TU activation pattern for $\mathbf{s}$ gives 
${\lfloor\log_2{n_{tu}\choose n_{rf}}\rfloor}$ MBM-TU index bits,
$ii)$ the non-zero entries in $\mathbf{s}$ gives $n_{rf}\log_2{M}$ 
QAM/PSK bits, and $iii)$ for each non-zero location $j$ in $\mathbf{s}$, 
$l_{j}$ gives $m_{rf}$ mirror index bits; since $\mathbf{s}$ has 
$n_{rf}$ non-zero entries, a total of $n_{rf}m_{rf}$ mirror index
bits are obtained from $l_j$'s.

\subsection{Average BEP analysis} 
The ML decision rule for GSM-MBM is given by (\ref{syseq2}).
The conditional pairwise error probability (PEP) of $\mathbf x$
being decoded as $\tilde{\mathbf{x}}$ can be written as
\begin{eqnarray}
P\left(\mathbf x\rightarrow\tilde{\mathbf{x}}|\mathbf{H}\right) = P\left( \|\vy - \mathbf{Hx}\|^2> \|\vy - \mathbf{H}\tilde{\mathbf{x}}\|^2|\mathbf{H}\right).
\label{syseq3}
\end{eqnarray}
From (\ref{syseq1}), we can write (\ref{syseq3}) as 
\begin{eqnarray}
P\left(\mathbf x\rightarrow\tilde{\mathbf{x}}|\mathbf{H}\right) & \hspace{-2mm} =& \hspace{-2mm}P\left( \|\vn\|^2> \| \mathbf{H}\left(\mathbf{x}-\tilde{\mathbf{x}}\right)+\vn\|^2|\mathbf{H}\right)\nonumber\\
& \hspace{-30mm} = & \hspace{-17mm} P\big(2\Re\big({\vn}^{\dagger}\mathbf{H}\left(\tilde{\mathbf{x}}-\mathbf{x}\right)\big)>\| \mathbf{H}\left(\mathbf{x}-\tilde{\mathbf{x}}\right)\|^2|\mathbf{H}\big),
\label{syseq4}
\end{eqnarray}
where $\Re{\left(.\right)}$ denotes real part, $\left(.\right)^\dagger$ 
denotes conjugate transpose, and 
$2\Re\left({\vn}^{\dagger}\mathbf{H}\left(\tilde{\mathbf{x}}-\mathbf{x}\right)\right)$ 
is a Gaussian random variable with zero mean and variance 
$2\sigma^2\| \mathbf{H}\left(\mathbf{x}-\tilde{\mathbf{x}}\right)\|^2$.
Therefore,
\begin{eqnarray}
P\left(\mathbf x\rightarrow\tilde{\mathbf{x}}|\mathbf{H}\right)&=& Q\Big(\sqrt{\| \mathbf{H}\left(\mathbf{x}-\tilde{\mathbf{x}}\right)\|^2/2\sigma^2} \Big).
\label{syseq5}
\end{eqnarray}
The computation of the unconditional PEP 
$P\left(\mathbf x\rightarrow\tilde{\mathbf{x}}\right)$ requires the 
expectation of the $Q\left(.\right)$ function in $(\ref{syseq5})$
w.r.t. $\mathbf H$, which can be obtained as follows \cite{alouini}:
\begin{eqnarray}
\hspace{-4mm}P\left(\mathbf x\rightarrow\tilde{\mathbf{x}}\right)\hspace{-3mm}&=&\hspace{-3mm}\mathbb{E}_{\mathbf H}\left\lbrace P\left(\mathbf x\rightarrow\tilde{\mathbf{x}}|\mathbf{H}\right)\right\rbrace\nonumber\\
&=&\hspace{-3mm}\mathbb{E}_{\mathbf H}\left\lbrace Q\Big(\sqrt{\| \mathbf{H}\left(\mathbf{x}-\tilde{\mathbf{x}}\right)\|^2/2\sigma^2}\Big)\right\rbrace \nonumber\\
&=&\hspace{-3mm} f\left(\beta\right)^{n_r}\sum_{i=0}^{n_r-1} {n_r-1+i\choose i}\left(1-f\left(\beta\right)\right)^i,
\end{eqnarray}
where 
$f\left(\beta\right) = \frac{1}{2}\left(1-\sqrt{\frac{\beta}{1+\beta}}\right)$ 
and $\beta = \frac{\|\mathbf{x}-\tilde{\mathbf{x}}\|^2}{4\sigma^2}$.
Now, an upper bound on the average BEP for GSM-MBM based on union bounding 
can be obtained as
\begin{eqnarray}
P_B \hspace{-1mm} &\leq & \hspace{-1mm} \frac{1}{2^\eta}\sum_{\mathbf{x}}\sum_{\tilde{\mathbf{x}}\neq\mathbf{x}} P\left(\mathbf x\rightarrow\tilde{\mathbf{x}}\right)\frac{\delta\left(\mathbf{x},\tilde{\mathbf{x}}\right)}{\eta},
\end{eqnarray}
where $\delta\left(\mathbf{x},\tilde{\mathbf{x}}\right)$ is the number of 
bits in which $\mathbf{x}$ differs from $\tilde{\mathbf{x}}$. The BEP upper 
bound for SIMO-MBM can be obtained from the above expression by setting 
$n_{tu}=n_{rf}=1$. Likewise, the BEP upper bound for MIMO-MBM can be 
obtained by setting $n_{tu}>1$ and $n_{rf}=n_{tu}$. The bit 
error rate (BER) performance of GSM-MBM and other multi-antenna 
schemes are presented in Sec. \ref{sec5a}. Note that the SIMO-MBM, 
MIMO-MBM, and GSM-MBM schemes do not need any feedback for their 
operation. In the next two sections, we study how feedback based 
physical layer techniques can be beneficial when used in MBM schemes.

\section{ED-based MAP selection}
\label{sec3}
In practice, an MBM-TU may be designed to have more RF mirrors 
available for use than the number of RF mirrors actually used.
Let $M_{rf}$ denote the number of mirrors available in an 
MBM-TU\footnote[2]{A typical implementation of an MBM-TU reported in 
the literature has $M_{rf}=14$ RF mirrors available \cite{mbm3}.}.
This means that the maximum number of channel fade symbols (aka
MBM constellation points) that can be generated by the MBM-TU is 
$2^{M_{rf}}$, i.e., one MBM constellation point per MAP. But not 
all $2^{M_{rf}}$ MAPs, and hence not all the corresponding MBM 
constellation points may be used. Only a subset of the $2^{M_{rf}}$ 
MAPs, say $2^{m_{rf}}$, $m_{rf} \leq M_{rf}$, MAPs are actually used 
to convey $m_{rf}$ bits through indexing mirrors. Now, one can choose 
the best subset of $2^{m_{rf}}$ MAPs from the set of all $2^{M_{rf}}$ 
possible MAPs. In other words, select the best $2^{m_{rf}}$ among the 
$2^{M_{rf}}$ MBM constellation points and form the MBM alphabet 
${\mathbb H}$ of size $|{\mathbb H}|=2^{m_{rf}}$ constellation 
points. Such MAP selection in MBM can be viewed as analogous to 
transmit antenna selection (TAS) in multi-antenna systems, where a 
subset of antennas among the available antennas is selected for 
transmission and the selection is based on channel knowledge. 

Here, we consider MAP selection in MIMO-MBM. Let 
$\mathbb{S}_{\scriptsize\mbox{all}}$ denote the set of all possible 
MAPs per MBM-TU. So, $|{\mathbb S}_{\scriptsize\mbox{all}}|=2^{M_{rf}}$. 
Let $\mathbb{S}_{\scriptsize\mbox{sub}}$ denote a possible subset of 
$\mathbb{S}_{\scriptsize\mbox{all}}$, where 
$|{\mathbb S}_{\scriptsize\mbox{sub}}|=2^{m_{rf}}$, $m_{rf}\leq M_{rf}$. 
The receiver estimates all the $|{\mathbb S}_{\scriptsize\mbox{all}}|$ 
MBM constellation points for every coherence interval, selects the best 
$|{\mathbb S}_{\scriptsize\mbox{sub}}|$ constellation points among them, 
and conveys the indices of the corresponding MAPs to the transmitter. 
The transmitter uses these selected MAPs to index the mirrors in that 
coherence interval. 

We consider two MAP selection schemes, one that uses an MI-based metric 
(which is studied in \cite{mbm2},\cite{mbm2a}) and another that uses an 
ED-based metric. ED-based antenna selection has been studied in the 
context of SM systems in \cite{sm_ed}. In \cite{yang_asm}, 
an ED-based adaptive spatial modulation (ASM) scheme has been presented, 
in which the receiver feeds back the most suitable modulation alphabet 
size for each of the transmit antennas based on the channel estimates 
for every coherence interval. Also, various link adaptation techniques 
like transmit precoding and antenna selection based on ED has been 
summarized in \cite{yang_design}, which also summarizes various SM 
variants like GSSK, GSM, and space-time shift keying (STSK). In what 
follows in this section, we present MI-based and ED-based MAP selection 
schemes. 

\subsection{MI-based MAP selection}
In \cite{mbm2},\cite{mbm2a}, a selection scheme that chooses the MBM 
constellation points with the highest energies is studied. This scheme 
is motivated by the observation that mutual information is  proportional 
to the energy (norm) of the MBM constellation point, and hence selecting 
the MBM constellation points with the highest energies maximizes the mutual 
information. 

Let $\mathbb{L}^j_{\tiny \mbox {MI}}=\{l_{j1},l_{j2},\cdots,l_{j{|\mathbb{S}_{\scriptsize\mbox{sub}}|}}\}$
be the set of MAP indices corresponding to the 
$|\mathbb{S}_{\scriptsize\mbox{sub}}|$ largest energies for the $j$th 
MBM-TU, which expects that
\[
\|{\mathbf{h}}_{l_{j1}}^j\|^2\geq\|{\mathbf{h}}_{l_{j2}}^j\|^2\geq\cdots\geq\|{\mathbf{h}}_{l_{j{|\mathbb{S}_{\tiny\mbox{sub}}|}}}^j\|^2\geq \cdots\geq\|{\mathbf{h}}_{l_{j{|\mathbb{S}_{\tiny\mbox{all}}|}}}^j\|^2,
\]
$j=1,2,\cdots,n_{t_u}$. Let 
$\mathbb{L}_{\tiny \mbox {MI}}=\{\mathbb{L}^1_{\tiny \mbox {MI}}, \mathbb{L}^2_{\tiny \mbox {MI}}, \hdots, \mathbb{L}^{n_{tu}}_{\tiny \mbox {MI}}\}$ 
be the set of MAP indices corresponding to all the MBM-TUs. The received 
signal vector considering this MAP selection is  given by
\begin{eqnarray}
\mathbf{y}=  {\mathbf H}_{{\tiny\mbox{MI}}} \mathbf{x}+\mathbf{n},
\end{eqnarray}
where 
${\mathbf H}_{\tiny\mbox{MI}}$ is the channel matrix of size 
$n_r\times n_{tu}|{\mathbb S}_{\scriptsize\mbox{sub}}|$, given by
${\mathbf H}_{\tiny\mbox{MI}}=[{\mathbf H}_{\tiny\mbox{MI}}^1 \ {\mathbf H}_{\tiny\mbox{MI}}^2 \ \cdots \ {\mathbf H}_{\tiny\mbox{MI}}^{n_{tu}}]$, and
${\mathbf H}_{\tiny\mbox{MI}}^j=[{\mathbf{h}}_{l_{j1}}^j \ {\mathbf{h}}_{l_{j2}}^j \ \cdots \ {\mathbf{h}}_{l_{j{|\mathbb{S}_{\tiny\mbox{sub}}|}}}^j ]$.
That is, the selected MBM vector constellation points of all the MBM-TUs
form the column vectors of the ${\mathbf H}_{\tiny\mbox{MI}}$ matrix. At 
the receiver, ML detection is performed using the knowledge of the channel 
matrix ${\mathbf H}_{\tiny\mbox{MI}}$.

\subsection{ED-based MAP selection}
Another way to do MAP selection is to choose the best MBM constellation 
points based on ED. Let $\mathcal{I}^j$ denote the collection of sets of 
MAP indices corresponding to the enumerations of the
${|\mathbb{S}_{\scriptsize\mbox{all}}|\choose {|\mathbb{S}_{\scriptsize\mbox{sub}}|}}$ 
combinations of selecting $|\mathbb{S}_{\scriptsize\mbox{sub}}|$ out of 
$|\mathbb{S}_{\scriptsize\mbox{all}}|$  MAPs of the $j$th MBM-TU. Let 
$\mathcal{L}$ denote the following set, defined as 
\begin{eqnarray}
\mathcal{L} \hspace{-3mm}&=&\hspace{-3mm}\big \{ \mathbb{L}=\{\mathbb{L}^1, \mathbb{L}^2,\cdots, \mathbb{L}^{n_{tu}}\}: \mathbb{L}^j \in \mathcal{I}^j, j=\{1, \cdots, n_{tu}\}\big\}. \nonumber
\end{eqnarray}
Among the $|\mathcal{L}|$ possible sets, choose that set which maximizes 
the minimum Euclidean distance among all possible transmit vectors. That is,
\begin{eqnarray}
{\mathbb L}_{{\tiny \mbox{ED}}}={\argmax_{{\mathbb L}\in \mathcal{L}}} \Big\{ \min_{\substack{\mathbf{x},\tilde{\mathbf{x}} \in \mathcal{X}\\ {\mathbf{x}}\neq {\tilde{\mathbf{x}}}}}||{\mathbf H}_{{\mathbb L}}\left(\mathbf{x}-{\tilde{\mathbf{x}}}\right)||^2\Big\},
\label{ed_map}
\end{eqnarray}
where $\mathbf{H}_{{\mathbb L}}$ is the channel matrix of size  
${n_r\times {n_{tu}}{|\mathbb{S}_{\scriptsize\mbox{sub}}|}}$ 
corresponding to the set $\mathbb{L}$, given by 
$\mathbf{H}_{{\mathbb L}} =\left[\mathbf{H}_{{\mathbb L}}^1 \hspace{1mm} \mathbf{H}_{{\mathbb L}}^2 \hspace{1mm} \cdots  \hspace{1mm}\mathbf{H}_{{\mathbb L}}^{n_{tu}}\right]$, 
${\mathbf H}_{\mathbb{L}}^j=[{\mathbf{h}}_{l_{j1}}^j \ {\mathbf{h}}_{l_{j2}}^j \ \cdots \ {\mathbf{h}}_{l_{j{|\mathbb{S}_{\tiny\mbox{sub}}|}}}^j ]$,  
$l_{j_k}$ is the $k$th element in the set $\mathbb{L}^j$, and $\mathcal{X}$ 
represents the set of all possible transmit vectors. The received signal 
vector considering this MAP selection is 
\begin{eqnarray}
\mathbf y &=& {\mathbf H}_{{\mathbb L}_{{\tiny\mbox{ED}}}}\mathbf x+\mathbf n.
\end{eqnarray}
At the receiver, ML detection is performed using the knowledge of the 
channel matrix ${\mathbf H}_{\mathbb{L}_{{\tiny \mbox{ED}}}}$. 
In Sec. \ref{sec5b}, we present the BER performance of the
ED-based and MI-based selection schemes.

{\em Complexity:}
The order of complexity of MI-based MAP selection scheme per MBM-TU is
$\mathcal{O}\left(|\mathbb{S}_{\scriptsize\mbox{all}}|\right)$. Hence,
the order of total complexity is
$\mathcal{O}\left(n_{tu}|\mathbb{S}_{\scriptsize\mbox{all}}|\right)$,
which is linear in $n_{tu}$ and  $|\mathbb{S}_{\scriptsize\mbox{all}}|$.
Whereas, the order of complexity of ED-based MAP selection in
$\left(\ref{ed_map}\right)$ is
$\mathcal{O} \left(|\mathcal{L}||\mathcal{X}|^2\right)$, where
$|\mathcal{L}|={{|\mathbb{S}_{\scriptsize\mbox{all}}|}\choose {|\mathbb{S}_{\scriptsize\mbox{sub}}|}}^{n_{tu}}$
and
$|\mathcal{X}|={\left({|\mathbb{S}_{\scriptsize\mbox{sub}}|}M\right)}^{n_{tu}}$,
which is exponential in $n_{tu}$ and $|\mathbb{S}_{\scriptsize\mbox{all}}|$.
Note that complexity in ED-based selection  is dependent on $M$, whereas
MI-based selection scheme complexity does not depend on $M$. The set
$\mathcal{L}$ is assumed to be known to both the transmitter and the
receiver. If $\mathbb{L}_{{\tiny \mbox{MI}}}$ in MI based scheme (or
$\mathbb{L}_{{\tiny \mbox{ED}}}$ in ED based scheme) is the $k$th element
in the set $\mathcal{L}$ (i.e., $k$ is the index of the selected set), then
the receiver feeds back the integer value $k$ to the transmitter. Since
$|\mathcal{L}|={{|\mathbb{S}_{\scriptsize\mbox{all}}|}\choose {|\mathbb{S}_{\scriptsize\mbox{sub}}|}}^{n_{tu}}$,
the number of feedback bits required is
${\big\lceil\log_2{{{|\mathbb{S}_{\scriptsize\mbox{all}}|}\choose {|\mathbb{S}_{\scriptsize\mbox{sub}}|}}^{n_{tu}}}\big\rceil}$.

\subsubsection{Diversity analysis}
In this subsection, we present an analysis of the diversity order 
achieved by the ED-based MAP selection scheme. 
We can write 
$\mathbf{H}_{\mathbb{L}}$ as
$\mathbf{H}_{\mathbb{L}}=\mathbf{H}\mathbf{A}_{\mathbb{L}}$, where  
$\mathbf{H}$ is the channel matrix of size  
${n_r\times {n_{tu}}{|\mathbb{S}_{\scriptsize\mbox{all}}|}}$, given by 
$\mathbf{H} =\left[\mathbf{H}^1 \hspace{1mm} \mathbf{H}^2 \hspace{1mm} \cdots  \hspace{1mm}\mathbf{H}^{n_{tu}}\right]$, 
${\mathbf H}^j=[{\mathbf{h}}_{1}^j \ {\mathbf{h}}_{2}^j \ \cdots \ {\mathbf{h}}_{|\mathbb{S}_{\tiny\mbox{all}}|}^j ]$, 
$\mathbf{A}_{\mathbb{L}}$ is the MAP selection matrix of size  
${{n_{tu}}{|\mathbb{S}_{\scriptsize\mbox{all}}|}\times {n_{tu}}{|\mathbb{S}_{\scriptsize\mbox{sub}}|}}$ 
corresponding to the set $\mathbb{L}$, given by
$\mathbf{A}_{\mathbb{L}}=\text{diag}\{\mathbf{A}_{\mathbb{L}^1}, \mathbf{A}_{\mathbb{L}^2}, \cdots, \mathbf{A}_{\mathbb{L}^{n_{tu}}}\}$,
and 
$\mathbf{A}_{\mathbb{L}^j}=[\mathbf{e}_{l_{j1}} \mathbf{e}_{l_{j2}} \cdots \mathbf{e}_{l_{j|\mathbb{S}_{\tiny\mbox{sub}}|}}]$. 
Note that for every $j$, $\mathbf{A}_{\mathbb{L}^j}$  can have at most 
one non-zero element in each row and each column. Now, (\ref{ed_map}) can 
be written as 
\begin{eqnarray}
{\mathbb L}_{{\tiny \mbox{ED}}} &=& {\argmax_{{\mathbb L}\in \mathcal{L}}} \Big\{ \min_{\substack{\mathbf{x},{\tilde{\mathbf{x}}} \in \mathcal{X}\\ {\mathbf{x}}\neq {\tilde{\mathbf{x}}}}}||{\mathbf H}\mathbf{A}_{{\mathbb L}}\left(\mathbf{x}-{\tilde{\mathbf{x}}}\right)||^2\Big\} \nonumber\\
&=& {\argmax_{{\mathbb L}\in \mathcal{L}}} \Big\{ \min_{\substack{\mathbf{z} ,\tilde{\mathbf{z}}\in\mathcal{X}_\mathbb{L}\\{\mathbf{z}}\neq {\tilde{\mathbf{z}}}}}||{\mathbf H}\left(\mathbf{z}-{\tilde{\mathbf{z}}}\right)||^2\Big\},
\end{eqnarray}
where $\mathcal{X}_\mathbb{L}$ is the set corresponding to $\mathbb{L}$ 
defined as 
$\mathcal{X}_\mathbb{L}=\{\mathbf{z}: \mathbf{z}=\mathbf{A}_\mathbb{L}\mathbf{x}, \mathbf{x} \in \mathcal{X}\}$. 
Let $\triangle \mathcal{X}_{\mathbb{L}}$ be the set of difference vectors 
corresponding to the set $\mathcal{X}_{\mathbb{L}}$, i.e., 
$\triangle\mathcal{X}_{\mathbb{L}}
=\{\mathbf{z}-{\tilde{\mathbf{z}}}:\mathbf{z}, {\tilde{\mathbf{z}}}\in\mathcal{X}_{\mathbb{L}}, \mathbf{z} \neq {\tilde{\mathbf{z}}}\}$. 
Let $\triangle\mathcal{D}$ be the set of matrices defined as
\begin{eqnarray}
\triangle\mathcal{D}\hspace{-0.5mm}= \hspace{-0.5mm}\left\{\mathbf{D}=[\mathbf{d}_1 \hspace{0.5mm} \mathbf{d}_2  \hspace{0.5mm} \cdots \hspace{0.5mm}  \mathbf{d}_{|\mathcal{L}|}]:\mathbf{d}_k \in \triangle\mathcal{X}_{\mathbb{L}_k},  k=1, \cdots, |\mathcal{L}|\right\}, \nonumber
\end{eqnarray}
where $\mathbb{L}_k$ is the $k$th element in the set $\mathcal{L}$. The 
size of each matrix in $\triangle\mathcal{D}$ is 
$n_{tu}|\mathbb{S}_{\scriptsize\mbox{all}}| \times |\mathcal{L}|$. 
The following proposition gives the diversity order achieved by the
ED-based MAP selection scheme in MIMO-MBM.

{\bf Proposition 1.}
{\em The diversity order achieved by the ED-based MAP selection scheme
in MIMO-MBM is given by}
$d = n_r\left(|\mathbb{S}_{\scriptsize\mbox{all}}|-|\mathbb{S}_{\scriptsize\mbox{\mbox sub}}|+1\right)$.

{\em Proof:} The proof is given in Appendix A.

In Sec. \ref{sec5b}, we present the numerical results that
validate this Proposition. It can be shown that the diversity order 
achieved by ED-based MAP selection in GSM-MBM is also given by $d$. 
It is further noted that the diversity order for ED-based 
TAS for SM and V-BLAST systems with ML detection has been analyzed in 
\cite{sm_edas} and \cite{tas_vblast_diver}, respectively, where it has 
been shown that the diversity order in both these systems is given by 
$n_r(n_t-n_s+1)$, where $n_t$ is the total number of available transmit 
antennas and $n_s$ is the number of selected antennas in TAS.

\section{Phase compensation and constellation rotation}
\label{sec4}
In this section, we study the performance of another feedback based 
transmission scheme called the {\em phase compensation and constellation 
rotation (PC-CR) scheme}. A PC-CR scheme in the context of generalized 
SSK has been studied in \cite{pc1}. This scheme exploits the knowledge 
of the random channel phases (not the amplitudes) at the transmitter to 
enhance performance. The idea is to co-phase the channels of the 
active transmit antennas for any spatial-constellation point. That is, 
the channel phases are compensated at the transmitter, which can be 
viewed as equal-gain combining (EGC) at the transmitter using knowledge 
of channel phases at the transmitter. The co-phased spatial-constellation 
points are further phase-rotated by a deterministic angle which is chosen
from $[0, 2\pi)$, so that the minimum Euclidean distance of the 
constellation points at the receiver is maximized. Here, we study the 
performance of the PC-CR scheme applied to MIMO-MBM. Consider $n_{tu}$ 
MBM-TUs at the transmitter, where each MBM-TU uses $m_{rf}$ mirrors. 
Assume that each MBM-TU transmits a tone.

\subsection{Case of $n_r=1$}
\label{sec4a}
Consider the case when $n_r=1$. The number of MAPs is $N_m=2^{m_{rf}}$.
For every coherence interval, the receiver estimates all the MBM 
constellation points, i.e., estimates $h_{1,k}^{j}$ for every 
$k \in \{1,2,\cdots,N_m\}$, $j \in \{1,2,\cdots,n_{tu}\}$. Let 
$|h_{1,k}^j|$ and $\phi_{1,k}^j$ denote the magnitude and phase of 
$h_{1,k}^j$. The receiver feeds back all the phases, i.e., 
$\phi_{1,k}^{j}$ for every 
$k \in \{1,2,\cdots,N_m\}$, $j \in \{1,2,\cdots,n_{tu}\}$, to the 
transmitter. Assume that the feedback is perfect. Using this feedback, 
the transmitter co-phases (i.e., phase compensates) the channel 
corresponding to the active MAP in each MBM-TU. Specifically, let
$\mathbf{u}$ denote the phase-compensated transmit vector obtained by 
multiplying the transmit vector $\mathbf{x}$ by phase compensation 
matrix, given by
$\mathbf{W}=\text{diag}\{[(\boldsymbol{\phi}_1^1)^T \ (\boldsymbol{\phi}_1^2 )^T \ \cdots  (\boldsymbol{\phi}_1^{n_{tu}})^T ]\}$, 
where 
$\boldsymbol{\phi}_1^j=[e^{-\imath\phi_{1,1}^j}\ e^{-\imath\phi_{1,2}^j} \ \cdots \ e^{-\imath\phi_{1,N_m}^j}]^T$, $j\in \{1, 2, \cdots, n_{tu}\}$, 
and $\imath=\sqrt{-1}$.

Let 
$\mathbb{U}_{\scriptsize\mbox{pc}} \Define \{\mathbf{u}: \mathbf{u}=\mathbf{W}\mathbf{x}, \mathbf{x}\in \mathcal{X}\}$,
denote the phase-compensated signal set, where $\mathcal{X}$ represents 
the set of all possible transmit vectors without phase compensation. 
After phase compensation, the resultant phase-compensated transmit 
vectors are further rotated to improve performance. Specifically,
denoting the $k$th vector in $\mathbb{U}_{\scriptsize\mbox{pc}}$ as 
$\mathbf{u}_k$, each element in $\mathbf{u}_k$ is rotated by the angle 
$\psi_k$. The rotation angles $\{\psi_k\}_{k=1}^{|\mathcal{X}|}$ are 
chosen such that the minimum Euclidean distance of the constellation at 
the receiver is maximized. The optimum angles are obtained as the
solution to the following optimization problem:
\begin{eqnarray}
\{\hat{\psi_k}\}\hspace{-1mm}=\hspace{-1mm}{\argmax_{\substack{{\psi_k}\in [0,2\pi)\\ \ \forall k }}}\hspace{-1mm}\Big\{\min_{\substack{\mathbf{u}_{k_1}, \mathbf{u}_{k_2}\in {\mathbb{U}_{\scriptsize\mbox{pc}}}\\{k_1}\neq{k_2}}}\hspace{-2mm}\|{\tilde{\mathbf h}}_1^T\hspace{-1mm}\left(\mathbf{u}_{k_1}e^{\imath{\psi_{k_1}}}-\mathbf{u}_{k_2}e^{\imath{\psi_{k_2}}}\right)\|^2\Big\}, \nonumber
\end{eqnarray}
where 
$\tilde{\mathbf{h}}_1\hspace{-0mm}=\hspace{-0mm}
[h_{1,1}^1 \cdots h_{1,N_m}^1 h_{1,1}^2 \cdots h_{1,N_m}^2\hspace{-1mm} \cdots  h_{1,1}^{n_{tu}} \cdots h_{1,N_m}^{n_{tu}}]^T$.
Taking a geometrical view of the above optimization problem, we can 
see that its solution is given by $\hat{\psi_k}=(k-1)2\pi/|\mathcal{X}|$. 

Let $\mathbb{V}_{\scriptsize\mbox{pc-cr}}$ denote the resulting signal 
set after phase compensation and constellation rotation described above. 
The $k$th vector in $\mathbb{V}_{\scriptsize\mbox{pc-cr}}$, denoted by
$\mathbf{v}_k$, is then given by $e^{\imath \hat{\psi_k}}\mathbf{u}_k$.
The received signal at the receiver can be written as 
\begin{equation}
y = \tilde{\mathbf{h}}_1^T\mathbf{v}+n, 
\label{syseq_pccr}
\end{equation}
and the corresponding ML decision rule is given by
\begin{equation}
\hat{\vv}= \argmin_{\vv \in {\mathbb V}_{\scriptsize\mbox{pc-cr}}} |y-\tilde{\mathbf{h}}_1^T\mathbf{v}|^2.
\label{syseq_pccr_ml}
\end{equation}
Now, from $\hat{\mathbf{v}}$, the detected $\vx$ vector, denoted by
$\hat\vx$, can be obtained as $\hat\vx=(\hat\vv^{\dagger})^T\odot \hat\vv$,
where $\odot$ denotes the element-wise multiplication operator.
The $\hat{\mathbf{x}}$ vector is demapped to get the corresponding 
information bits. 

\subsection{Case of $n_r > 1$}
\label{sec4b}
When there are more than one receive antenna, the phase compensation 
presented in the previous subsection for $n_r=1$ is not directly 
applicable, since there are $n_r>1$ complex-valued channels between 
each MBM-TU and the receiver. Let
$\tilde{\mathbf{h}}_k=
[h_{k,1}^1\ \cdots\ h_{k,N_m}^1 \ h_{k,1}^2\ \cdots\ h_{k,N_m}^2 \cdots \ h_{k,1}^{n_{tu}} \ \cdots \ h_{k,N_m}^{n_{tu}}]^T$
denote the channel coefficient vector of size $N_mn_{tu}\times 1$
of the $k$th receive antenna, $k=1,2,\cdots,n_r$. Here, we present two 
receiver schemes for phase compensation when $n_r>1$.

\subsubsection{Receiver scheme 1}
\label{rx. scheme 1}
A possible extension of phase compensation for multiple receive antennas 
($n_r > 1$) is presented in \cite{pc2}, in which the upper bound on the 
conditional BEP for the ML decision rule in (\ref{syseq_pccr_ml}) is 
evaluated for each receive antenna and the receive antenna with the
lowest upper bound is selected. We refer this scheme as receiver 
scheme 1 (Rx. scheme 1). The upper bound on the conditional BEP (i.e., 
given $\tilde{\mathbf{h}}_k$)  for ML detection in (\ref{syseq_pccr_ml}) 
is given by
\begin{eqnarray}
P_{B|\tilde{\mathbf{h}}_k} \hspace{-3mm}&\leq &\hspace{-3mm} \frac{1}{2^\eta}\sum_{\mathbf{v}_1}\sum_{\mathbf{v}_2\neq\mathbf{v}_1} P\left({\mathbf v}_1\rightarrow{\mathbf{v}_2}|\tilde{\mathbf{h}}_k\right)\frac{\delta\left(\mathbf{v}_1,{\mathbf{v}_2}\right)}{\eta}\nonumber\\
&\hspace{-20mm} =&\hspace{-12mm} \frac{1}{2^\eta}\sum_{\mathbf{v}_1}\sum_{{\mathbf{v}_2}\neq\mathbf{v}_1} \hspace{-0mm}Q\bigg(\sqrt{\frac{|\tilde{\mathbf{h}}_k^T\left(\mathbf{v}_1-{\mathbf{v}_2}\right)|^2}{2\sigma^2}}\bigg)\frac{\delta\left(\mathbf{v}_1,{\mathbf{v}_2}\right)}{\eta}.
\label{pccr_ml_upper}
\end{eqnarray}
The receiver selects the receive antenna with lowest upper bound, i.e., 
\begin{eqnarray}
\hat{k} = \argmin_{k \in \{1, 2 , \cdots, n_r\}} P_{B|\tilde{\mathbf{h}}_k}.
\label{rx_scheme_sel}
\end{eqnarray}
The receiver feeds back all the phases of $\tilde{\mathbf{h}}_{\hat{k}}$. 
Let $\mathbb{V}_{\scriptsize\mbox{pc-cr}}^{\hat{k}}$ denote signal set
corresponding to the phase compensation and constellation rotation 
defined as in Sec. \ref{sec4a}. Only the selected receive antenna (i.e., 
$\hat{k}$) will be active and others will be silent. The received 
signal can then be written as
\begin{equation}
y = \tilde{\mathbf{h}}_{\hat{k}}^T\mathbf{v}+n,
\label{pccr_2}
\end{equation}
and the corresponding ML decision rule is given by
\begin{equation}
\hat{\vv}=\argmin_{{\vv} \in {\mathbb V}_{\scriptsize\mbox{pc-cr}}^{\hat{k}}}
|y - \tilde{\mathbf{h}}_{\hat{k}}^T\mathbf{v}|^2.
\label{pccr_ml_2}
\end{equation}
Now, from $\hat{\mathbf{v}}$, the detected $\vx$ vector, denoted by
$\hat\vx$, can be obtained as $\hat\vx=(\hat\vv^{\dagger})^T\odot \hat\vv$.
A drawback in this scheme is that it uses only one antenna to receive
signal even though multiple antennas are available at the receiver. To 
overcome this drawback, we present another possible extension of phase 
compensation scheme for multiple receive antennas, referred as receiver 
scheme 2 (Rx. scheme 2) in which signals from all the receive antennas 
will be used for detection.

\subsubsection{Receiver scheme 2}
\label{rx_scheme 2}
The receiver selects the receive antenna  for phase compensation as in 
Sec. \ref{rx. scheme 1}, and feeds back its corresponding phases to the 
transmitter. Let $\hat{k}$ denote the selected receive antenna, and let 
$\mathbb{V}_{\scriptsize\mbox{pc-cr}}^{\hat{k}}$ denote signal set 
corresponding to the phase compensation and constellation rotation. 
The signals from all the receive antennas are used. Then, the received 
signal vector is given by
\begin{equation}
\mathbf{y} = \mathbf{H}\mathbf{v}+\mathbf{n},
\label{syseq_pccr_2}
\end{equation}
where $\mathbf{H}$ is $n_r\times N_mn_{tu}$ channel matrix given by 
$\mathbf{H}=[\tilde{\mathbf{h}}_1 \ \tilde{\mathbf{h}}_2 \ \cdots \ \tilde{\mathbf{h}}_{n_r}]^T$. 
Since phase compensation is carried out based on the phases of 
$\hat{k}$th receive antenna, the effect of phase compensation needs to 
be eliminated at other receive antennas. To account for this, we present 
the modified decision rule as follows: 
\begin{equation}
\hat{\vv}=\argmin_{{\vv} \in {\mathbb V}_{\scriptsize\mbox{pc-cr}}^{\hat{k}}}
\|\mathbf{y} - \mathbf{H}^{(\mathbf{v})}\mathbf{v}\|^2,
\label{syseq_pccr_ml_2}
\end{equation}
where 
$\mathbf{H}^{(\mathbf{v})}\triangleq[\tilde{\mathbf{h}}_1^{(\mathbf{v})} \cdots \ \tilde{{\mathbf{h}}}_{\hat{k}}^{(\mathbf{v})} \cdots \ \tilde{\mathbf{h}}_{n_r}^{(\mathbf{v})}]^T$, 
and $\tilde{\mathbf{h}}_i^{(\mathbf{v})}$'s are given by
\begin{eqnarray}
\tilde{\mathbf{h}}_i^{(\mathbf{v})} \triangleq \left\{ \begin{array}{ll}
\tilde{\mathbf{h}}_i  & \mbox{if } i = \hat{k} \\
\tilde{\mathbf{h}}_i \odot (\mathbf{v}^{\dagger})^T & \mbox{if } i\neq \hat{k}.
\end{array}
\right.
\end{eqnarray}
Since $\mathbf{x}=(\hat\vv^{\dagger})^T\odot \hat\vv$, we have 
$(\tilde{\mathbf{h}}_i^{(\mathbf{v})})^T\mathbf{v}=\tilde{\mathbf{h}}_i^T\mathbf{x}$ 
for $i\neq {\hat{k}}$. Now, (\ref{syseq_pccr_ml_2}) becomes
\begin{eqnarray}
\hat{\vv} &\hspace{-1mm}=& \hspace{-1mm} {\argmin_{{\vv} \in {\mathbb V}_{\scriptsize\mbox{pc-cr}}^{\hat{k}}}} |y_{\hat{k}} - \tilde{\mathbf{h}}_{\hat{k}}^T\mathbf{v}|^2+\sum_{i\neq {\hat{k}}} |y_i - \tilde{\mathbf{h}}_i^T{((\hat\vv^{\dagger})^T\odot \hat\vv)}|^2\nonumber\\
&\hspace{-1mm}=& \hspace{-1mm} {\argmin_{{\vv} \in {\mathbb V}_{\scriptsize\mbox{pc-cr}}^{\hat{k}}}} {|y_{\hat{k}} - \tilde{\mathbf{h}}_{\hat{k}}^T\mathbf{v}|^2}+\sum_{i\neq {\hat{k}}} {|y_i - \tilde{\mathbf{h}}_i^T{\mathbf{x}}|^2}.
\label{syseq_pccr_ml_2_modified}
\end{eqnarray}
From (\ref{syseq_pccr_ml_2_modified}), we can see that this decision 
rule gives the advantage of both phase compensation (by $\hat{k}$th 
receive antenna) and SNR gain by using other $n_r-1$ receive 
antennas. 

\subsubsection{Diversity analysis}
\label{subsec_div_anal}
In this subsection, we present an analysis of the diversity order 
achieved by the PC-CR scheme. We present the analysis considering 
the case of $n_r>1$ with Rx. scheme 1 at the receiver. A similar
analysis applies to Rx. scheme 2 as well as the scheme with $n_r=1$.

The conditional BEP $P_{B|\tilde{\mathbf{h}}_k}$ in (\ref{pccr_ml_upper}) 
can be approximated by applying the nearest neighbor approximation in the 
high SNR region (\cite{gold_smith}, Eq. 5.45), as
\begin{eqnarray}
\hspace{-5mm}P_{B|\tilde{\mathbf{h}}_k} &\hspace{-3mm}\approx \hspace{-3mm}& Q\left(\sqrt{\frac{1}{2\sigma^2}{\min\limits_{{{\mathbf v_1}\neq {\mathbf v_2}}}|\tilde{\mathbf{h}}_k^T\left(\mathbf{v}_1-{\mathbf{v}_2}\right)|^2}}\right)\frac{\delta(\mathbf v_1, \mathbf v_2)}{\eta}.
\label{pccr_ml_upper_a}
\end{eqnarray}
Let $\tilde{\mathbf{h}}=[\tilde{\mathbf{h}}_1^T\ \tilde{\mathbf{h}}_2^T\ \cdots 
\ \tilde{\mathbf{h}}_{n_r}^T]^T$ 
denote the channel coefficient vector of size $N_mn_{tu}n_r \times 1$. Then, 
we can write $\tilde{\mathbf{h}}_k^T=\tilde{\mathbf{h}}^T \mathbf{B}_k$, 
where $\mathbf B_k$ is the receive antenna selection matrix of size 
$n_rn_{tu}N_m \times n_{tu}N_m$ corresponding to the $k$th receive antenna, 
which is given by 
$\mathbf{B}_k=[\mathbf{e}_{(k-1)n_{tu}N_m+1}\hspace{1mm} \mathbf{e}_{(k-1)n_{tu} N_m+2} \hspace{1mm} \cdots \hspace{1mm} \mathbf{e}_{kn_{tu}N_m}].$
Now, (\ref{rx_scheme_sel}) can be written as
\begin{eqnarray}
\hat{k}\hspace{-1mm} &=&\hspace{-1mm} {\argmax_{{k}\in \{1,\cdots,n_r\}}} \bigg \{ \min_{\substack{\mathbf{v}_1,\mathbf{v}_2 \in {\mathbb V}_{\scriptsize\mbox{pc-cr}}^{{k}}\\ {\mathbf{v}_1}\neq {\mathbf{v}_2}}}|\tilde{\mathbf{h}}_{{k}}^T\left(\mathbf{v}_1-\mathbf{v}_2\right)|^2\bigg\} \nonumber\\
\hspace{-1mm}&=& \hspace{-1mm} {\argmax_{{k}\in \{1,\cdots,n_r\}}} \bigg\{ \min_{\substack{\mathbf{v}_1,\mathbf{v}_2 \in {\mathbb V}_{\scriptsize\mbox{pc-cr}}^{{k}}\\ {\mathbf{v}_1}\neq {\mathbf{v}_2}}}|\tilde{\mathbf{h}}^T\mathbf{B}_k\left(\mathbf{v}_1-\mathbf{v}_2\right)|^2\bigg\}\hspace{-0.0mm}. 
\label{opt_pccr}
\end{eqnarray}
The following proposition gives the diversity order 
achieved by the PC-CR scheme.

{\bf Proposition 2.}
{\em The diversity order achieved by the PC-CR scheme 
is given by} $d_{\scriptsize{\mbox{pc-cr}}}= n_r\left(n_{tu}+1\right)$. 

{\em Proof:} The proof is given in Appendix B.

In Sec. \ref{sec5c}, we present the numerical results that
validate this Proposition. Section \ref{sec5c} also presents the BER
performance of schemes without and with PC-CR.

\section{Results and discussions}
\label{sec5}
The numerical results and discussions for the GSM-MBM scheme (in Sec. 
\ref{sec2}), the MAP selection schemes (in Sec. \ref{sec3}), and the 
PC-CR scheme (in Sec. \ref{sec4}) are presented in this section in 
Secs. \ref{sec5a}, \ref{sec5b}, and \ref{sec5c}, respectively. 

\subsection{Performance of GSM-MBM} 
\label{sec5a}
\subsubsection{Comparison between systems with and without RF mirrors}
\begin{figure}
\centering
\includegraphics[height=2.3in,width=3.4in]{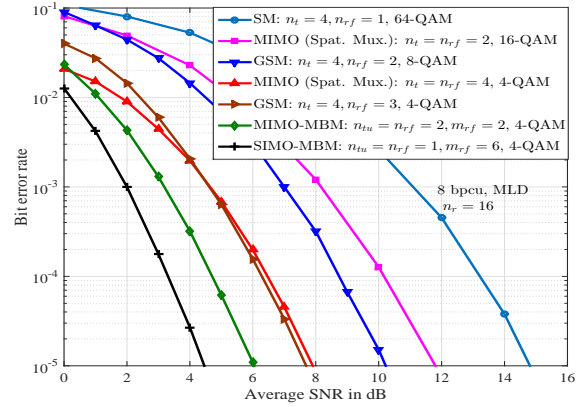}
\vspace{-2mm}
\caption{BER performance comparison between MBM schemes with 
RF mirrors (SIMO-MBM and MIMO-MBM) and other multi-antenna schemes without 
RF mirrors (MIMO, SM, GSM) at 8 bpcu and $n_r=16$.}
\label{perf0}
\vspace{-4mm}
\end{figure}

First, in Fig. \ref{perf0}, we illustrate the effectiveness of MBM 
schemes with RF mirrors compared to other popularly known multi-antenna 
schemes without RF mirrors. The spectral efficiency is fixed at 8 bpcu for 
all the schemes considered. All the schemes use $n_r=16$ and ML detection. 
The schemes considered are: $i)$ SIMO-MBM with $n_{tu}=n_{rf}=1$, $m_{rf}=6$, 
and 4 QAM (6 bits from indexing RF mirrors and 2 bits from one 4-QAM symbol), 
$ii)$ MIMO-MBM with $n_{tu}=n_{rf}=2$, $m_{rf}=2$, and 4 QAM (4 bits from 
indexing RF mirrors and 4 bits from two 4-QAM symbols), $iii)$ MIMO (spatial 
multiplexing) with $n_{t}=2$, $n_{rf}=2$, and 16-QAM (8 bits from two 16-QAM 
symbols), $iv)$ MIMO (spatial multiplexing) with $n_{t}=4$, $n_{rf}=4$,
and 4-QAM (8 bits from four 4-QAM symbols), $v)$ SM with $n_t=4$, $n_{rf}=1$, 
and 64-QAM (2 bits from indexing antennas and 6 bits from one 64-QAM symbol),
$vi)$ GSM with $n_t=4$, $n_{rf}=2$, and 8-QAM (2 bits from indexing antennas 
and 6 bits from two 8-QAM symbols), and $vii)$ GSM with $n_t=4$, $n_{rf}=3$, 
and 4-QAM (2 bits from indexing antennas and 6 bits from three 4-QAM symbols).
Note that among the above schemes, SIMO-MBM and MIMO-MBM 
are schemes which use RF mirrors and the others are non-MBM schemes 
which do not use RF mirrors. It can be seen that the MBM schemes (i.e., 
SIMO-MBM and MIMO-MBM) with RF mirrors achieve better BER 
performance compared to other multi-antenna schemes which do not use RF 
mirrors. The SIMO-MBM and MIMO-MBM schemes perform better than non-MBM 
schemes because of the use of RF mirror index bits, small QAM size (4-QAM), 
and no interference ($n_{rf}=1$ in SIMO-MBM)/less interference ($n_{rf}=2$ 
in MIMO-MBM). This illustrates the BER performance advantage possible with 
systems that employ media-based modulation using RF mirrors. 
Note that both SIMO-MBM and MIMO-MBM in this example use 4-QAM to achieve
8 bpcu. In this case, SIMO-MBM performs better than MIMO-MBM because 
there is no spatial interference in SIMO-MBM whereas there is spatial
interference in MIMO-MBM. Further note that, since the number of RF 
mirrors used is more in SIMO-MBM ($m_{rf}=6$) than in MIMO-MBM ($m_{rf}=2$),
the number of MAPs given by $n_{tu}2^{m_{rf}}$ is more in SIMO-MBM 
($1\times 2^{6}=64$) than in MIMO-MBM ($2\times 2^2=8$). This means 64 
pilot channel uses are needed in SIMO-MBM, whereas only 8 pilot channel 
uses are needed in MIMO-MBM.
\begin{figure}
\centering
\includegraphics[height=2.3in,width=3.4in]{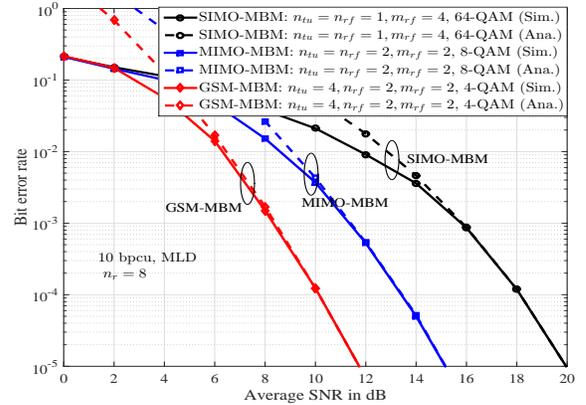}
\vspace{-2mm}
\caption{BER performance of SIMO-MBM, MIMO-MBM, and GSM-MBM 
with $n_r=8$, and 10 bpcu. SIMO-MBM: $n_{tu}=n_{rf}=1$, $m_{rf}=4$, 
64-QAM. MIMO-MBM: $n_{tu}=n_{rf}=2$, $m_{rf}=2$, 8-QAM. GSM-MBM: 
$n_{tu}=4$, $n_{rf}=2$, $m_{rf}=2$, 4-QAM.}
\label{perf1}
\vspace{-4mm}
\end{figure}

\subsubsection{Comparison between SIMO-MBM, MIMO-MBM, GSM-MBM}
Next, we evaluate the BER performance of GSM-MBM scheme through 
analysis and simulations. We also evaluate the bit error performance 
of SIMO-MBM and MIMO-MBM schemes for comparison. We compare these three 
schemes for the same spectral efficiency. Figure \ref{perf1} shows the 
BER performance comparison between the following schemes, namely SIMO-MBM, 
MIMO-MBM, and GSM-MBM schemes, all achieving the same 10 bpcu: $i)$ SIMO-MBM 
using $n_{tu}=n_{rf}=1$, $m_{rf}=4$, and 64-QAM (4 bits from indexing mirrors 
and 6 bits from one 64-QAM symbol), $ii)$ MIMO-MBM using $n_{tu}=n_{rf}=2$, 
$m_{rf}=2$, and 8-QAM (4 bits from indexing mirrors, 6 bits from two 8-QAM 
symbols), and $iii)$ GSM-MBM using $n_{tu}=4$, $n_{rf}=2$, $m_{rf}=2$, and 
4-QAM (4 bits from indexing mirrors, 2 bits from indexing MBM-TUs, and 
4 bits from two 4-QAM symbols).
All the three schemes use $n_r=8$ and ML detection. The following 
observations can be made from Fig. \ref{perf1}. The analytical upper 
bound is tight for moderate-to-high SNRs. It is seen 
that MIMO-MBM achieves better performance compared to SIMO-MBM. For 
example, at a BER of $10^{-4}$, MIMO-MBM requires about 4.4 dB less 
SNR compared to SIMO-MBM. This is because, although MIMO-MBM has
spatial interference, it has the benefit of using a lower QAM
size compared to SIMO-MBM (8-QAM in MIMO-MBM and 64-QAM in SIMO-MBM).
GSM-MBM is found to perform better than both SIMO-MBM and MIMO-MBM. 
For example, at $10^{-4}$ BER, GSM-MBM gives an SNR advantage of about 
3.2 dB and 7.8 dB over MIMO-MBM and SIMO-MBM, respectively. This is 
because more bits are conveyed through indexing in GSM-MBM (i.e., 
through indexing of mirrors and MBM-TUs), which results in a reduced 
QAM size (4-QAM for GSM-MBM compared to 8-QAM and 64-QAM for MIMO-MBM 
and SIMO-MBM, respectively). The results, therefore, show that GSM is 
an attractive physical layer technique which can be beneficial when 
used in MBM.

\subsubsection{Effect of spatial correlation}
In the analysis and simulation results presented above,
the $h_{i,k}^j$s are considered to be i.i.d. However, due to space
limitation in the MBM-TU, there can be spatial correlation effects. 
For example, the channel fades corresponding to different MAPs (i.e., 
corresponding to the different ON/OFF status of the RF mirrors) in an 
MBM-TU can be correlated. Likewise, the fades corresponding to the MAPs 
of different MBM-TUs can also be correlated. Here, we study the effect 
of these correlations on the performance of MBM. We also present trellis 
coded modulation (TCM) based symbol mapping to alleviate these correlation 
effects.

We use the Kronecker model \cite{tcsm}, which is commonly used 
to model spatial correlation. The correlated channel matrix in the 
Kronecker model is given by 
\begin{eqnarray}
\mathbf{H}=\mathbf{R}_{\tiny \mbox{Rx}}^{1/2}\tilde{\mathbf{H}}\mathbf{R}_{\tiny \mbox{Tx}}^{1/2},
\label{kronecker}
\end{eqnarray}
where $\mathbf{R}_{\tiny \mbox{Rx}}$ is the $n_r\times n_r$ receive 
correlation
matrix, $\tilde{\mathbf{H}}$ is a  
matrix of size
$n_r\times N_mn_{tu}$ whose entries are i.i.d. and distributed as
$\mathcal{CN}(0,1)$, and $\mathbf{R}_{\tiny \mbox{Tx}}$ is the
$N_mn_{tu}\times N_mn_{tu}$ transmit correlation matrix. In our system, 
the transmit correlation matrix is determined by two types of correlation, 
one among fades across MBM-TUs and another across MAPs in an MBM-TU. The 
exponentially decaying correlation model \cite{Lokya} is used to characterize 
the correlation between MBM-TUs, and the equi-correlated model is used to 
characterize the correlation between fades across MAPs in an MBM-TU. 
Accordingly, the transmit correlation matrix $\mathbf{R}_{\tiny \mbox{Tx}}$ 
is written as 
\begin{eqnarray*}
\mathbf{R}_{\tiny \mbox{Tx}} = \begin{bmatrix}
\mathbf{R}_{1,1} & \mathbf{R}_{1,2} & \cdots & \mathbf{R}_{1,n_{tu}} \\
\mathbf{R}_{2,1} & \mathbf{R}_{2,2} & \cdots & \mathbf{R}_{2,n_{tu}} \\
\vdots & \vdots & \ddots & \vdots \\
\mathbf{R}_{n_{tu},1} & \mathbf{R}_{n_{tu},2} & \cdots & \mathbf{R}_{n_{tu},n_{tu}}
\end{bmatrix},
\end{eqnarray*}
where $\mathbf{R}_{i,j}$ is an $N_m \times N_m$ matrix whose $(k,l)$th 
entry is the correlation coefficient between $k$th MAP of the $i$th 
MBM-TU and $l$th MAP of the $j$th MBM-TU. Note that $i=j$ corresponds 
to correlations across MAPs in an MBM-TU, and $i\neq j$ corresponds to 
correlations across MBM-TUs. Let $\rho_m$ denote the correlation 
coefficient in the equi-correlation model 
in an MBM-TU, i.e., the diagonal elements of 
$\mathbf{R}_{i,i}$ are 1 and the off-diagonal elements are $\rho_m$.
Let $\rho_a^{|i-j|}$ denote the correlation coefficient in the 
exponentially decaying correlation model 
across $i$th 
and $j$th MBM-TUs, i.e., $\mathbf{R}_{i,j}$ for $i\neq j$ is given by 
$\mathbf{R}_{i,j}=\rho_a^{|i-j|}\mathbf{1}$, where $\mathbf{1}$
represents all ones matrix of size $N_m\times N_m$. Based on the above, 
the $\mathbf{R}_{\tiny \mbox{Tx}}$ matrix for an example system 
with $n_{tu}=3$ and $m_{rf}=1$ is given by 
\begin{eqnarray*}
\mathbf{R}_{\tiny \mbox{Tx}}  =
\left[\begin{array}{c c : c c : c c}
1 & \rho_m & \rho_a & \rho_a & \rho_a^2 & \rho_a^2 \\
\rho_m & 1 &\rho_a & \rho_a &  \rho_a^2 & \rho_a^2 \\ \hdashline
\rho_a & \rho_a & 1 & \rho_m & \rho_a & \rho_a \\
\rho_a & \rho_a & \rho_m & 1 & \rho_a & \rho_a \\ \hdashline
\rho_a^2 & \rho_a^2 & \rho_a & \rho_a & 1 & \rho_m \\
\rho_a^2 & \rho_a^2 & \rho_a & \rho_a &\rho_m & 1
\end{array}\right].
\end{eqnarray*}
The receive correlation matrix $\mathbf{R}_{\tiny \mbox{Rx}}$ is
also considered to follow the exponentially decaying correlation model,
i.e., the $(i,j)$th entry of  $\mathbf{R}_{\tiny \mbox{Rx}}$ is given
by $\rho_a^{|i-j|}$. 

\begin{figure}
\centering
\includegraphics[height=2.3in,width=3.4in]{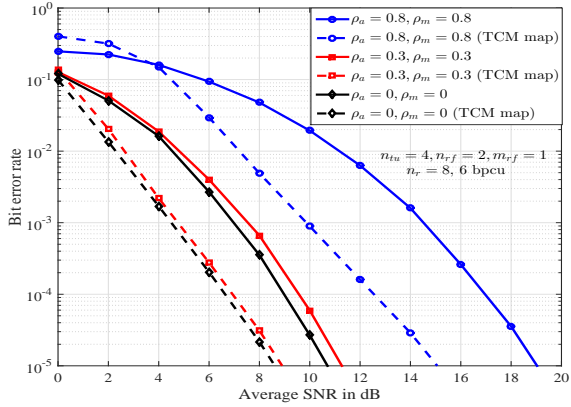}
\vspace{-2mm}
\caption{Effect of spatial correlation on the performance 
of GSM-MBM with $n_{tu}=4$, $n_{rf}=2$, $m_{rf}=1$, $n_r=8$, and 6 bpcu.
For system with no TCM encoding: BPSK. MLD. For system with TCM encoding: 
64-state trellis convolution encoder of rate 6/8, 4-QAM, soft Viterbi
decoding.}
\label{corr_fig}
\vspace{-4mm}
\end{figure}

In Fig. \ref{corr_fig}, we illustrate the effect of spatial correlation
on the BER performance of GSM-MBM with $n_{tu}=4$, $n_{rf}=2$, $m_{rf}=1$, 
BPSK, 6 bpcu, and $n_r=8$. It can be seen that, as would be 
expected, spatial correlation degrades the BER performance. Higher the 
correlation, more is the degradation in BER. For example, compared to the 
system with no correlation (i.e., $\rho_a=\rho_m=0$), the systems with 
$\rho_a=\rho_m=0.3$ and $\rho_a=\rho_m=0.8$ experience degradation of 
about 0.5 dB and 7.5 dB, respectively, at 
$10^{-3}$ BER. In addition to illustrating the performance degradation 
due to spatial correlation, this figure also illustrates how trellis 
coded modulation (TCM) can alleviate this degradation. 
In \cite{tcsm},\cite{code_design}, bit mapping techniques based on TCM 
have been proposed to improve the performance of SM systems in the 
presence of spatial correlation. In \cite{tcsm}, only the index bits 
are convolutionally encoded, whereas both the index and QAM/PSK bits 
are convolutionally encoded in \cite{code_design}. Here, we consider 
a TCM mapping scheme where all the bits are convolutionally encoded as 
in \cite{code_design} prior to GSM-MBM encoding. In the TCM coded
GSM-MBM scheme the rate is kept the same as 6 bpcu by 
using TCM with a 64-state trellis convolutional code of rate 6/8 and 
4-QAM. The rows of the $6\times 8$ octal generator matrix of the 
convolutional encoder used are [1 1 1 0 0 0 1 0], [3 1 2 0 0 0 0 0],
[2 5 5 0 0 0 0 0], [0 0 0 1 1 1 0 1], [0 0 0 3 1 2 0 0], and 
[0 0 0 2 5 5 0 0].
A frame size of 20 channel uses and soft decision 
Viterbi decoder are used. It can be seen in Fig. \ref{corr_fig} that, for
the same bpcu, TCM encoding improves the BER performance of GSM-MBM in the
presence of spatial correlation. For example, for $\rho_a=\rho_m=0.3$ and
$\rho_a=\rho_m=0.8$, TCM results in an improvement of about 2.5 
dB and 4.5 dB, respectively, at a BER of $10^{-3}$. We have
observed similar improvements for various combinations of $\rho_a$ and 
$\rho_m$ values. 

\subsection{Performance of MAP selection schemes}
\label{sec5b}
\begin{figure}
\centering 
\includegraphics[height=2.3in,width=3.4in]{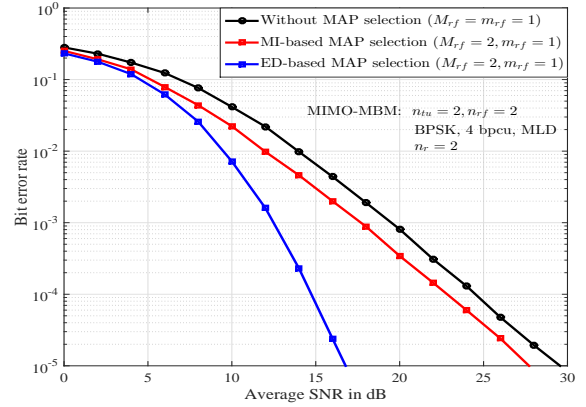}
\vspace{-2mm}
\caption{BER performance comparison between MIMO-MBM schemes without 
and with MAP selection, $n_{tu}=n_{rf}=2$, BPSK, 4 bpcu, and $n_r=2$: 
$i)$ no MAP selection with $M_{rf}=m_{rf}=1$, 
$ii)$ MI-based MAP selection with $M_{rf}=2$, $m_{rf}=1$, and
$iii)$ ED-based MAP selection with $M_{rf}=2$, $m_{rf}=1$.}
\label{perf3}
\vspace{-2mm}
\end{figure}

\begin{figure}
\centering
\subfigure[All constellation points]{
\includegraphics[height=1.20in]{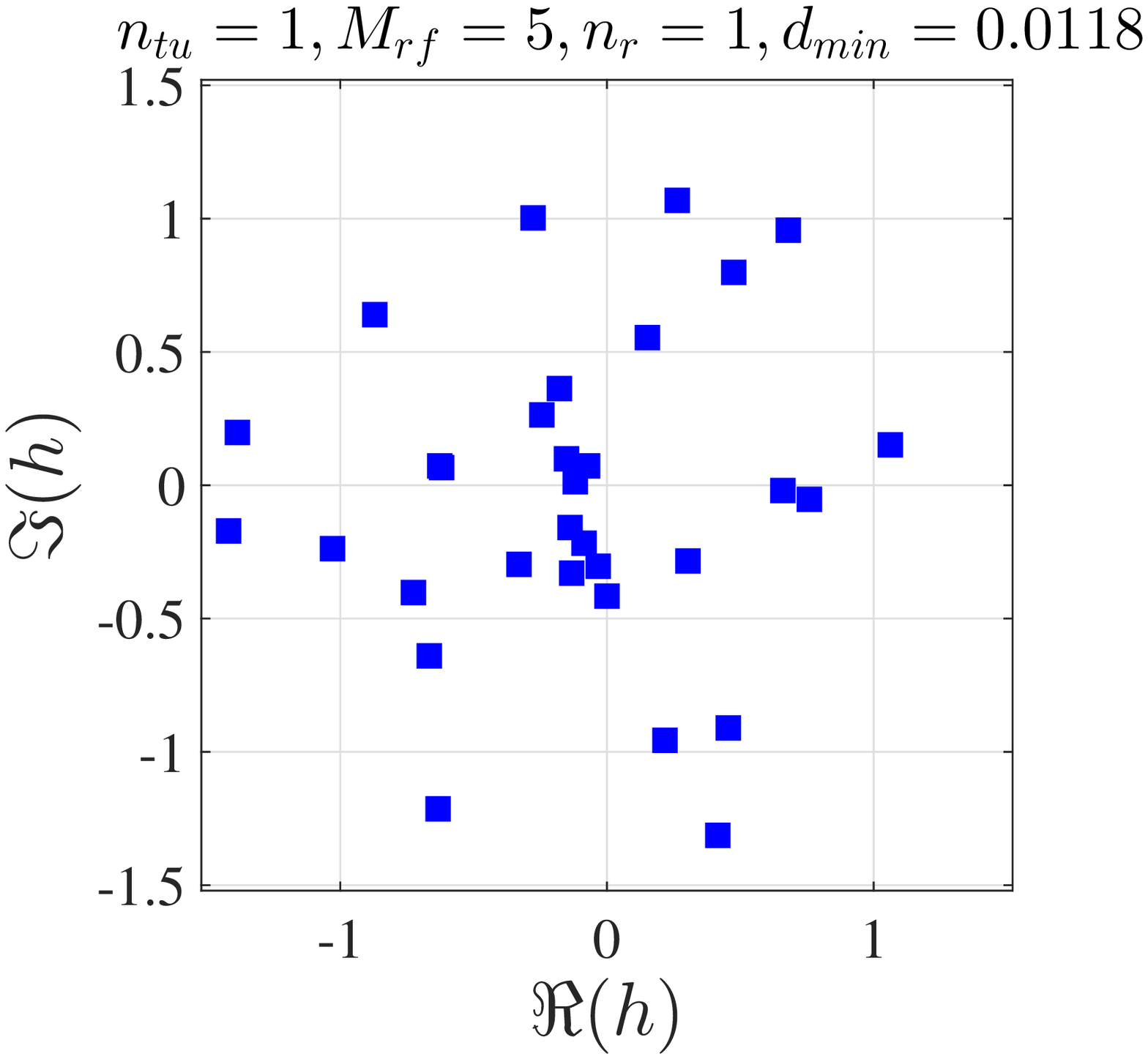}
\label{sel_a}
}
\hspace{2cm}
\subfigure[MI-based selection]{
\includegraphics[height=1.325in]{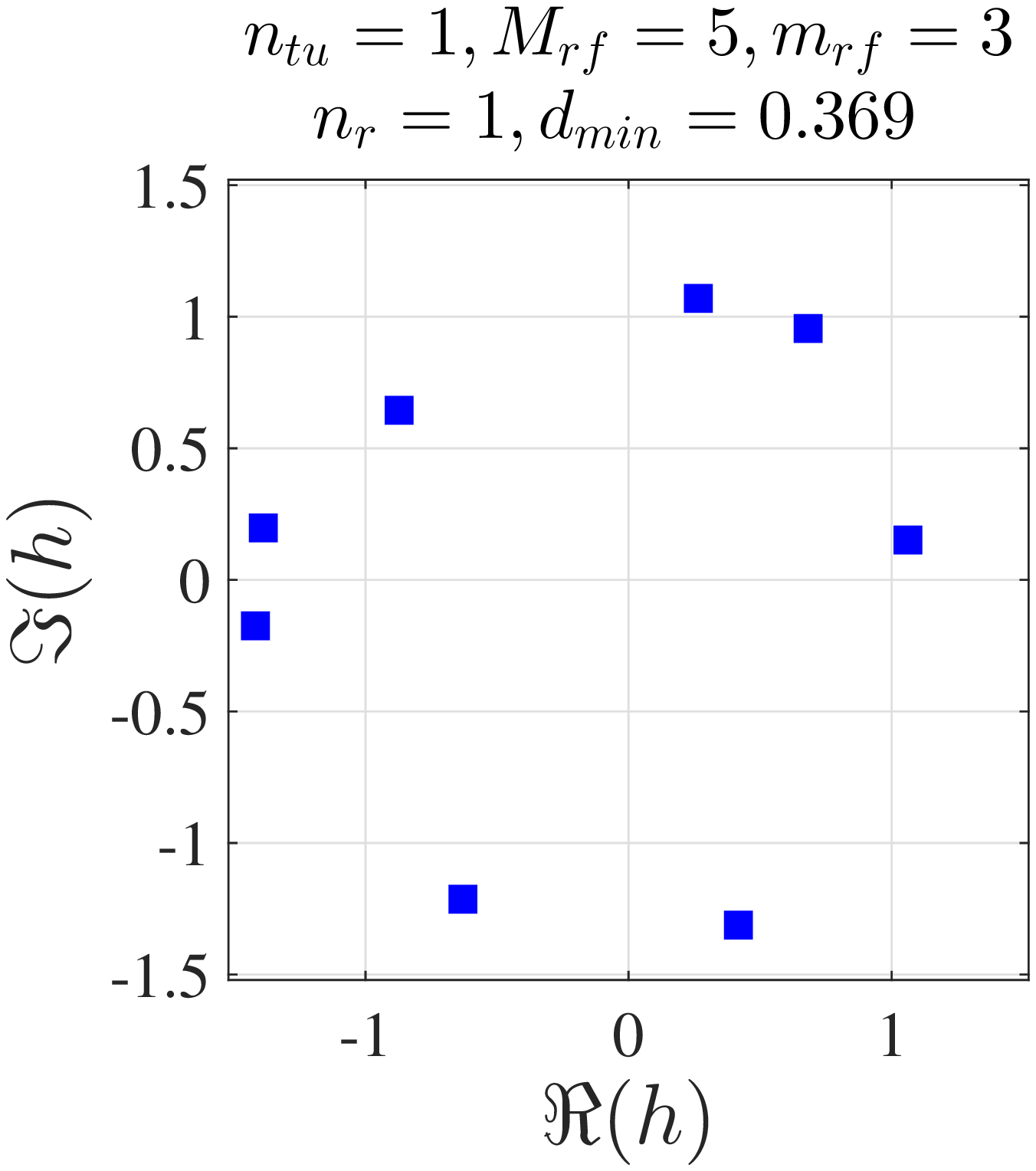}
\label{sel_b}
}
\hspace{-10mm}
\subfigure[ED-based selection]{
\includegraphics[height=1.325in]{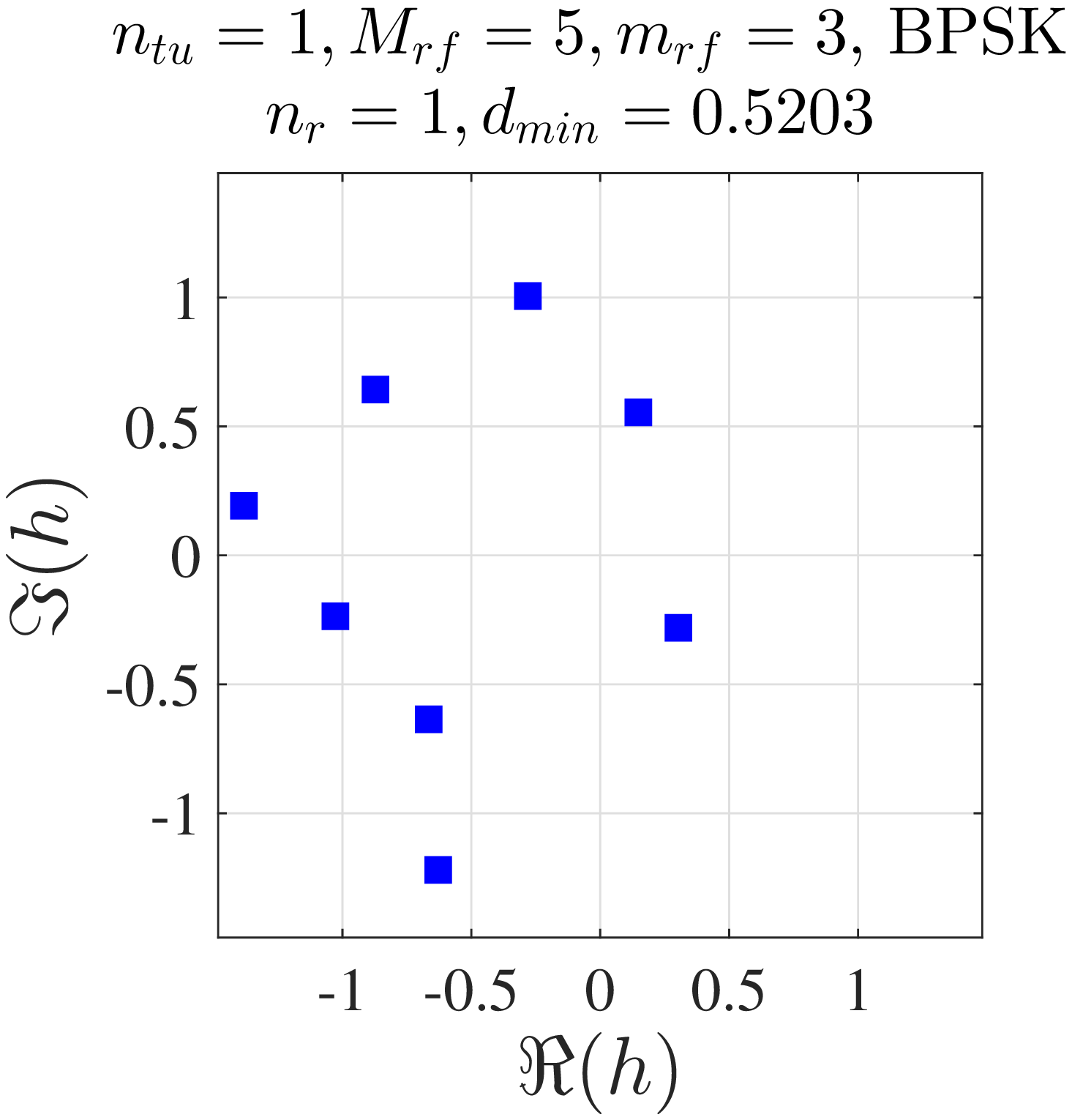}
\label{sel_b}
}
\caption{Constellation diagrams without and with MAP selection. 
(a) Set of all constellation points. 
(b) Constellation points selected by MI-based selection. 
(c) Constellation points selected by ED-based selection.}
\vspace{-5mm}
\label{sel_fig}
\end{figure}

In Fig. \ref{perf3}, we present a comparison between the BER performance
achieved by MIMO-MBM schemes without and with MAP selection. Three 
schemes, all with $n_{tu}=n_{rf}=2$, BPSK, 4 bpcu, and $n_r=2$, are 
considered. The first scheme is a scheme with no MAP selection, i.e.,
$M_{rf}=m_{rf}=1$. The second scheme is a scheme with MI-based MAP 
selection where $M_{rf}=2$ and $m_{rf}=1$. The third scheme is same
as the second scheme, except that MAP selection is done based on ED.  
In all the three schemes, two bits through BPSK symbols (one bit on each 
MBM-TU) and two bits through RF mirror indexing (one bit on each MBM-TU) 
result in 4 bpcu. As expected, we observe that the MAP selection
schemes achieve better performance compared to the scheme without 
MAP selection. This is because of the better minimum distance between 
constellation points achieved by the selection schemes. We also see that 
ED-based selection achieves significantly better performance compared to 
MI-based selection. In fact, ED-based selection achieves a higher 
diversity order compared to MI-based selection. Again, the reason for 
this is that, because it maximizes the minimum Euclidean distance, the 
constellation points chosen by ED-based selection have better minimum 
distance between them compared those chosen by MI-based selection. This 
can be observed in Fig. \ref{sel_fig}, which shows the constellation 
diagrams for the selection schemes with $n_{tu}=n_{rf}=n_r=1$, 
$M_{rf}=5,m_{rf}=3$, and BPSK. The minimum distance between the 
constellation points, $d_{min}$, are $0.0118$, $0.3690$, and $0.5263$ 
for the schemes without selection, MI-based selection, and ED-based 
selection, respectively. 

\begin{figure}
\centering 
\includegraphics[height=2.3in,width=3.4in]{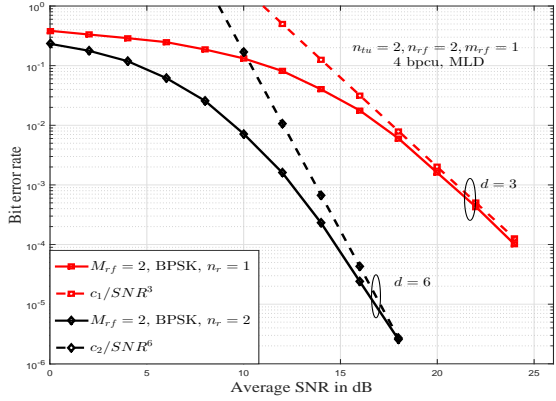}
\vspace{-2mm}
\caption{Diversity orders achieved by ED-based MAP selection in
MIMO-MBM for various system parameters.}
\label{perf4}
\vspace{-4mm}
\end{figure}

Figure \ref{perf4} presents a validation of the diversity orders of 
ED-based selection predicted by Proposition 1. The slopes of the 
simulated BER plots in the high SNR regime show that the achieved 
diversity orders are 3 and 6 for the schemes with $M_{rf}=2$, 
$m_{rf}=1$, and $n_r=1$ and 2, respectively, which are the same as 
the ones (i.e., $ n_r\left(2^{M_{rf}}-2^{m_{rf}}+1\right)$) 
proved analytically by Proposition 1. The constants 
used in Fig. \ref{perf4} are $c_1=2000$ and $c_2=169990$.

\begin{figure}
\centering 
\includegraphics[height=2.3in,width=3.4in]{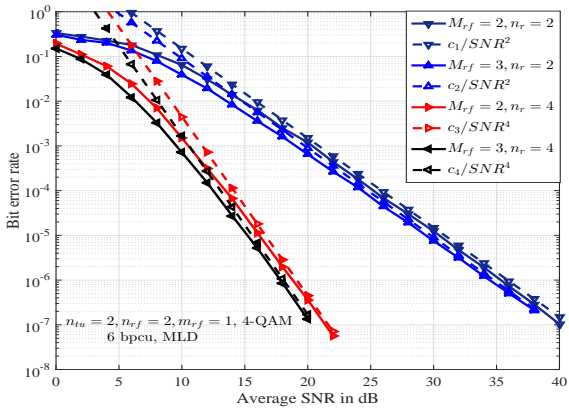}
\vspace{-2mm}
\caption{Diversity orders achieved by MI-based MAP selection 
in MIMO-MBM with $n_{tu}=n_{rf}=2, m_{rf}=1$, 4-QAM, and 6 bpcu for various 
values of $M_{rf}$ and $n_r$: $i)$  $M_{rf}=2, n_r=2$; 
$ii)$ $M_{rf}=3, n_r=2$; 
$iii)$ $M_{rf}=2, n_r=4$; 
$iv)$ $M_{rf}=3, n_r=4$.}
\label{perf_mi}
\vspace{-4mm}
\end{figure}

We carried out simulations to predict the diversity order of the 
MI-based selection scheme. The obtained results are shown in Fig. 
\ref{perf_mi}. From this figure, it is seen that the diversity order 
achieved is $n_r$. The reason for this can be explained as follows. 
The distance between any two MBM constellation points is a sum of 
$n_r$ independent random variables, and selecting the constellation 
points based on energy (i.e., MI) will not change the distance 
properties. Therefore, the diversity order of the PEPs in which the 
MAP indices are distinct is $n_r$, and hence the diversity order of 
MI-based MAP selection scheme is $n_r$. The constants used in 
Fig. \ref{perf_mi} are $c_1=15$, $c_2=9$, $c_3=45$, and $c_4=17$. 

\subsection{Performance of PC-CR scheme}
\label{sec5c} 
\begin{figure}
\centering 
\includegraphics[height=2.3in,width=3.4in]{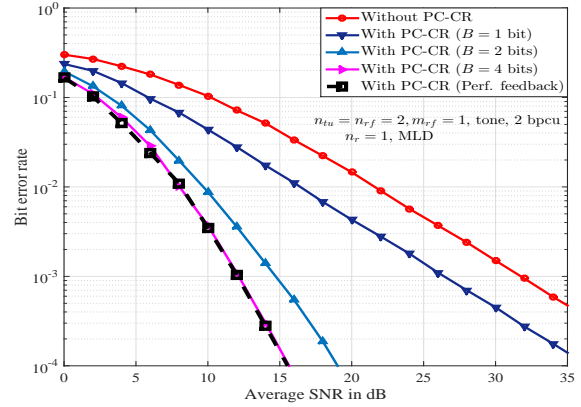}
\vspace{-2mm}
\caption{BER performance of MIMO-MBM without and with PC-CR 
(perfect feedback and limited feedback) 
for $n_{tu}=n_{rf}=2$, $m_{rf}=1$, tone, 2 bpcu, 
$n_r=1$, and ML detection.} 
\label{perf5}
\vspace{-4mm}
\end{figure}

Figure \ref{perf5} shows the BER performance of MIMO-MBM without and 
with PC-CR, for $n_{tu}=n_{rf}=2$, $m_{rf}=1$, tone, 2 bpcu, $n_r=1$, 
and ML detection. It can be seen that the feedback based PC-CR scheme 
significantly improves the BER performance; e.g., PC-CR 
scheme with perfect feedback is found to achieve an improved performance 
of about 20 dB at $10^{-3}$ BER compared to the scheme 
without PC-CR. This is because of the maximization of the minimum ED at 
the receiver in the PC-CR scheme.  
Note that the number of phase values to be fed back in the PC-CR scheme is
$n_{tu}2^{m_{rf}}$, which is exponential in $m_{rf}$. To study the effect 
of limited feedback on the performance of PC-CR, we consider that each 
feedback phase value is quantized using $B$ bits and these quantized 
bits are fed back. Since $n_{tu}2^{m_{rf}}$ phase values need to be 
fed back, the number of feedback bits required is  $n_{tu}2^{m_{rf}}B$. 
Since the phases are uniformly distributed in $(-\pi, \pi]$, the 
quantization levels are $-\pi+\frac{2\pi k}{2^B}$, $1 \leq k \leq 2^B$. 
For each feedback phase value, the receiver finds the nearest quantization 
level and feeds back its corresponding $k$ using $B$ bits to the transmitter. 
In Fig. \ref{perf5}, we illustrate the effect of number of feedback bits 
$B$ on the BER performance. It can be seen that the performance with 1-bit 
feedback is severely degraded, i.e., there is a degradation of about 14 dB 
at $10^{-3}$ BER compared to the case of perfect feedback. However, 
increasing $B$ from 1 bit to 2 bits significantly improves the performance
(by about 11 dB at $10^{-3}$ BER). It can be further noted that with just 
4-bit feedback ($B=4$), performance very close to perfect feedback is 
achieved.

\begin{figure}
\centering 
\includegraphics[height=2.3in,width=3.4in]{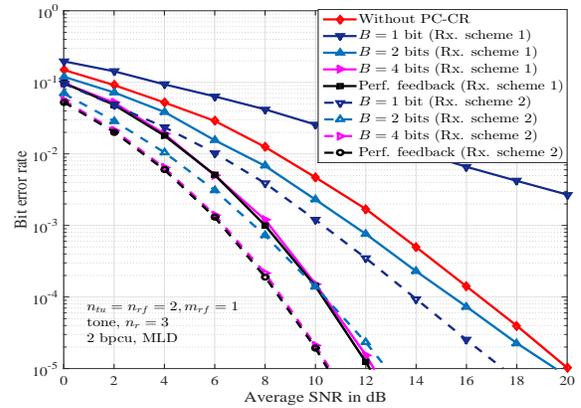}
\vspace{-2mm}
\caption{BER performance of MIMO-MBM without and with PC-CR 
using Rx. scheme 1 and Rx. scheme 2 for $n_{tu}=n_{rf}=2$, $m_{rf}=1$, 
tone, 2 bpcu, $n_r=3$, and ML detection.}
\vspace{-4mm}
\label{perf6}
\end{figure}

Figure \ref{perf6} shows the BER performance of MIMO-MBM 
without and with PC-CR using Rx. scheme 1 and Rx. scheme 2 for 
$n_{tu}=n_{rf}=2$, $m_{rf}=1$, tone, 2 bpcu, $n_r=3$, and ML detection.
This figure also shows the performance with limited feedback 
for $B=1,2,$ and 4 bits per feedback phase. It can be seen that $B=4$-bit 
feedback is sufficient to achieve performance very close to that with 
perfect phase feedback. Further it can be seen that the Rx.  
scheme 2 performs better than Rx. scheme 1 by about 1.5 dB at $10^{-5}$ 
BER. This is because Rx. scheme 2 uses signals from all the receive 
antennas for detection, whereas Rx. scheme 1 uses the signal only from 
the selected receive antenna.

\begin{figure}
\centering 
\includegraphics[height=2.3in,width=3.4in]{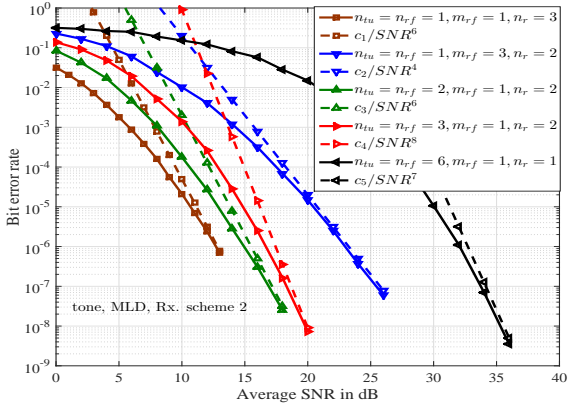}
\vspace{-2mm}
\caption{Diversity orders achieved by MIMO-MBM with PC-CR using 
Rx. scheme 2 for different system configurations: 
$i)$ $n_{tu}=1, m_{rf}=1, n_r=3$ (1 bpcu); 
$ii)$ $n_{tu}=2, m_{rf}=1, n_r=2$ (2 bpcu); 
$iii)$ $n_{tu}=1, m_{rf}=3, n_r=2$ (3 bpcu); 
$iv)$ $n_{tu}=3, m_{rf}=1, n_r=2$ (3 bpcu); 
$v)$ $n_{tu}=6, m_{rf}=1, n_r=1$ (6 bpcu).}
\label{pccr_eta1}
\vspace{-4mm}
\end{figure}

Figure \ref{pccr_eta1} presents a validation of the diversity 
orders predicted by Proposition 2. The slopes of the  simulated BER plots 
in the high SNR  regime show that the achieved diversity orders are 6, 6, 4, 
8, and 7 for the schemes with ($n_{tu}=1$, $m_{rf}=1$, $n_r=3$, 1 bpcu), 
($n_{tu}=2$, $m_{rf}=1$,  $n_r=2$, 2 bpcu), ($n_{tu}=1$, $m_{rf}=3$, $n_r=2$, 
3 bpcu), ($n_{tu}=3$, $m_{rf}=1$, $n_r=2$, 3 bpcu), and ($n_{tu}=6$, 
$m_{rf}=1$,  $n_r=1$, 6 bpcu), respectively, which are the same as the 
ones obtained from the diversity order given by $n_r(n_{tu}+1)$. 
The constants used in Fig. \ref{pccr_eta1} are $c_1=50, c_2=2000, 
c_3= 2000, c_4=9 \times 10^7$, and $c_5=4 \times 10^7$.

\section{Conclusions}
\label{sec6}
We investigated the performance of some interesting physical layer 
techniques when applied to media-based modulation (MBM), which is a 
recently proposed modulation scheme that uses RF mirrors to perturb the 
propagation environment to create independent channel fade realizations 
which themselves are used as the constellation points. The considered 
physical layer techniques included generalized spatial modulation (GSM), 
mirror activation pattern (MAP) selection (analogous to antenna selection 
in MIMO systems), and phase compensation and constellation rotation. It 
was shown that, for the same spectral efficiency, GSM-MBM can achieve 
better performance compared to MIMO-MBM. The Euclidean distance based 
MAP selection scheme was found to perform better than the mutual 
information based MAP selection scheme by several dBs. The diversity 
order achieved by the Euclidean distance based MAP selection scheme was 
shown to be $n_r(2^{M_{rf}}-2^{m_{rf}}+1)$, which was also validated
through simulations. Feedback based phase compensation and MBM 
constellation rotation was found to increase the Euclidean distance 
between the constellation points, thereby improving the bit error 
performance significantly. The diversity order achieved 
by the phase compensation and constellation rotation scheme was shown 
to be $n_r(n_{tu}+1)$, which was validated through simulations.

\section*{Appendix A}
\section*{Proof of Proposition 1} 
{\em Proof:}  
Let $d_{\mathbf{x},\tilde{\mathbf{x}}}$ denote the diversity order of 
PEP $P\left(\mathbf x\rightarrow\tilde{\mathbf{x}}\right)$. Then, the 
diversity order $(d)$ achieved by ED-based MAP selection scheme is 
given by
\begin{eqnarray}
d &=& \min_{\mathbf{x} \neq \tilde{\mathbf{x}}} d_{\mathbf{x},\tilde{\mathbf{x}}}.
\label{div_d}
\end{eqnarray}
Let ${\mathbb{L}_{{\tiny \mbox{ED}}}}$ be the 
$k_{\tiny \mbox{ED}}$th element in the set $\mathcal{L}$. In the following, 
we derive lower and upper bounds on $d$ and show that both these bounds 
turn out to be the same, given by
$n_r\left(|\mathbb{S}_{\scriptsize\mbox{all}}|-|\mathbb{S}_{\scriptsize\mbox{sub}}|+1\right)$.
  
\subsubsection{Lower bound on $d$}
The conditional PEP between $\mathbf{x}$ and $\tilde{\mathbf{x}}$ 
is given by
\begin{eqnarray}
P\left(\mathbf x\rightarrow\tilde{\mathbf{x}}|\mathbf{H}\right)&=&Q\hspace{-0mm}\big(\hspace{-0.5mm}\sqrt{\| \mathbf{H}\mathbf{A}_{\mathbb{L}_{{\tiny \mbox{ED}}}}\left(\mathbf{x}-\tilde{\mathbf{x}}\right)\|^2/{2\sigma^2}}\big) \nonumber\\
&\leq &\frac{1}{2}\text{exp}\big(-\| \mathbf{H}\left(\mathbf{z}-\tilde{\mathbf{z}}\right)\|^2/{4\sigma^2}\big),
\label{sys_prop1_1}
\end{eqnarray}
where $\mathbf{z}=\mathbf{A}_{\mathbb{L}_{{\tiny\mbox{ED}}}}\mathbf{x},  \tilde{\mathbf{z}}=\mathbf{A}_{\mathbb{L}_{{\tiny\mbox{ED}}}}{\tilde{\mathbf{x}}}$, 
$\mathbf{z},\tilde{\mathbf{z}}\in \mathcal{X}_{\mathbb{L}_{{\tiny\mbox{ED}}}}$,
and the inequality in (\ref{sys_prop1_1}) follows from Chernoff bound.
For a given  $k, k=1,\cdots,|\mathcal{L}|$, let 
$\mathbf{d}_{{\tiny \mbox{min}}}\left(k\right)$ represent the
difference vector in $\Delta \mathcal{X}_{\mathbb{L}_k}$ corresponding to 
the minimum ED, i.e., 
$\mathbf{d}_{\tiny \mbox{min}}\left(k\right)=\argmin_{d\in \Delta\mathcal{X}_{\mathbb{L}_k}}\|\mathbf{H}\mathbf{d}\|^2$. Let $\mathbf{D}_{\tiny\mbox{min}}$ 
is the matrix defined as 
$\mathbf{D}_{\tiny\mbox{min}}=\left[\mathbf{d}_{{\tiny \mbox{min}}}(1)\  \mathbf{d}_{{\tiny \mbox{min}}}(2)\ \cdots \ \mathbf{d}_{{\tiny \mbox{min}}}(|\mathcal{L}|)\right]$. 
Then, we have
\begin{eqnarray}
\hspace{-7mm}\| \mathbf{H}\left(\mathbf{z}-\tilde{\mathbf{z}}\right)\|^2 \hspace{-3mm}&\geq &\hspace{-3mm} \| \mathbf{H}\mathbf{d}_{\tiny \mbox{min}}\left(k_{\tiny \mbox{ED}}\right)\|^2 \label{sys_prop1_2}\\
&\geq &\hspace{-3mm}\frac{1}{|\mathcal{L}|}\| \mathbf{H}\mathbf{D}_{\tiny \mbox{min}}\|^2 \hspace{-0mm} = \hspace{-0mm} \frac{1}{|\mathcal{L}|}\text{Tr}\big(\mathbf{H}\mathbf{D}_{\tiny \mbox{min}}{\mathbf{D}_{\tiny \mbox{min}}}^\dagger {\mathbf{H}}^\dagger\big),
\label{sys_prop1_3}
\end{eqnarray}
where $\text{Tr}\left(.\right)$ denotes the trace operator, the inequality 
in (\ref{sys_prop1_2}) follows from the definition of 
$\mathbf{d}_{\tiny \mbox{min}}\left(k_{\tiny \mbox{ED}}\right)$, 
and the inequality in (\ref{sys_prop1_3}) follows from the fact
$\| \mathbf{H}\mathbf{d}_{\tiny \mbox{min}}\left(k_{\tiny \mbox{ED}}\right)\|^2 \geq \| \mathbf{H}\mathbf{d}_{\tiny \mbox{min}}\left(k\right)\|^2, 1 \leq k \neq k_{\tiny \mbox{ED}} \leq |\mathcal{L}|.$ 
Using eigenvalue decomposition, we have 
$\mathbf{D}_{\tiny \mbox{min}}{\mathbf{D}_{\tiny \mbox{min}}}^\dagger=\mathbf{U}\Lambda{\mathbf{U}}^\dagger$, 
where $\mathbf{U}$ is a unitary matrix and 
$\Lambda=\text{diag}\{\lambda_1,\lambda_2,\cdots,\lambda_p,0,0,\cdots,0\}$, 
$\lambda_i \neq 0$, $i=1,\cdots,p$, 
$p=rank\left(\mathbf{D}_{\tiny \mbox{min}}\right)$. Then, we have
\begin{eqnarray}
\hspace{-4mm}
\| \mathbf{H}\left(\mathbf{z}-\tilde{\mathbf{z}}\right)\|^2  \geq  
 \frac{1}{|\mathcal{L}|}\text{Tr}\big(\tilde{\mathbf{H}}\Lambda{\tilde{\mathbf{H}}}^\dagger\big) 
&\hspace{-3mm} = & \hspace{-3mm}\frac{1}{|\mathcal{L}|}\sum\limits_{i=1}^{n_r}\sum\limits_{j=1}^{p}\lambda_j |\tilde{h}_{i,j}|^2\nonumber\\
&\hspace{-3mm}\geq &\hspace{-3mm}\frac{\lambda_{\tiny\mbox{min}}}{|\mathcal{L}|} \sum\limits_{i=1}^{n_r}\sum\limits_{j=1}^{p} |\tilde{h}_{i,j}|^2,
\label{sys_prop1_4}
\end{eqnarray}
where 
$\tilde{\mathbf{H}}=\mathbf{H}\mathbf{U}$, $\lambda_{\tiny \mbox{min}}=\min_{\mathbf{D} \in \Delta\mathcal{D}} \lambda_s\left(\mathbf{D}{\mathbf{D}}^\dagger\right)$, 
and $\lambda_s\left(\mathbf{D}{\mathbf{D}}^\dagger\right)$ is the smallest 
non-zero eigenvalue of $\mathbf{D}{\mathbf{D}}^\dagger$. Since $\mathbf{U}$ 
is unitary, the entries of $\tilde{\mathbf{H}}$ are i.i.d. and 
$\mathcal{C}\mathcal{N}\left(0,1\right)$. From (\ref{sys_prop1_1}) 
and (\ref{sys_prop1_4}), we have
\begin{eqnarray}
P\left(\mathbf x\rightarrow\tilde{\mathbf{x}}|\mathbf{H}\right) \leq \frac{1}{2}\text{exp}\Bigg(-\frac{\lambda_{\tiny\mbox{min}}}{4\sigma^2|\mathcal{L}|}\bigg(\sum\limits_{i=1}^{n_r}\sum\limits_{j=1}^{p} |\tilde{h}_{i,j}|^2\bigg)\Bigg).
\end{eqnarray}
The unconditional PEP is then given by
\begin{eqnarray}
P\left(\mathbf x\rightarrow\tilde{\mathbf{x}}\right)\leq \mathbb{E}_{\mathbf H}\bigg\lbrace \frac{1}{2}\text{exp}\bigg(-\frac{\lambda_{\tiny\mbox{min}}}{4\sigma^2|\mathcal{L}|} \Big(\sum\limits_{i=1}^{n_r}\sum\limits_{j=1}^{p} |\tilde{h}_{i,j}|^2\Big)\bigg)\bigg\rbrace.
\label{sys_prop1_5}
\end{eqnarray}
Since $|\tilde{h}_{i,j}|^2$s are independent and exponentially distributed 
with unit mean, we have
\begin{eqnarray}
P\left(\mathbf x\rightarrow\tilde{\mathbf{x}}\right)\hspace{-1mm}&\leq &  \hspace{-1mm}\frac{1}{2}\prod\limits_{i=1}^{n_r}\prod\limits_{j=1}^{p}\left(1+\frac{\lambda_{\tiny\mbox{min}}}{4\sigma^2|\mathcal{L}|}\right)^{-1}. 
\end{eqnarray}
At high SNRs, $\frac{\lambda_{\tiny\mbox{min}}}{4\sigma^2|\mathcal{L}|}\gg 1$. 
Hence, we can write
\begin{eqnarray}
P\left(\mathbf x\rightarrow\tilde{\mathbf{x}}\right)\hspace{-1mm}&\leq &  \hspace{-1mm}\frac{1}{2}\Big(\frac{\lambda_{\tiny\mbox{min}}}{4\sigma^2|\mathcal{L}|}\Big)^{-n_rp}. 
\end{eqnarray}
Based on union bound, average BEP at high SNRs can be bounded as 
\begin{eqnarray}
P_B \hspace{-1mm} &\leq & \hspace{-1mm} \frac{1}{2^\eta}\sum_{\mathbf{x}}\sum_{\tilde{\mathbf{x}}\neq\mathbf{x}} P\left(\mathbf x\rightarrow\tilde{\mathbf{x}}\right)\frac{\delta\left(\mathbf{x},\tilde{\mathbf{x}}\right)}{\eta}\nonumber\\
&\leq & \frac{\left(2^\eta-1\right)}{2\eta} \Big(\frac{\lambda_{\tiny\mbox{min}}}{4\sigma^2|\mathcal{L}|}\Big)^{-n_rp},
\end{eqnarray}
which shows that the diversity order  achieved by ED based MAP selection 
scheme is lower bounded by $n_rp$. Next, we show that 
$p\geq \left(|\mathbb{S}_{\scriptsize\mbox{all}}|-|\mathbb{S}_{\scriptsize\mbox{sub}}|+1\right)$.
Any matrix $\mathbf{D} \in \triangle\mathcal{D}$ can be viewed in the form 
$\mathbf{D}=[\mathbf{D}_1^T\ \mathbf{D}_2^T\ \cdots\ \mathbf{D}_{n_{tu}}^T]^T$,
where $\mathbf{D}_j$ is a sub-matrix of size 
$|\mathbb{S}_{\scriptsize\mbox{all}}| \times |\mathcal{L}|$. Consider a 
matrix $\mathbf{D}\in \triangle\mathcal{D}$ which is constrained such 
that only one sub-matrix (say, $\mathbf{D}_k)$ is a non-zero sub-matrix 
and all other sub-matrices ($\mathbf{D}_j$'s$, j\neq k$) are zero 
sub-matrices. That is, the constrained matrix is of the form 
$\mathbf{D}=[ \mathbf{0}^T \ \mathbf{0}^T \ \cdots \ \mathbf{D}_k^T \ \cdots \ \mathbf{0}^T \ \mathbf{0}^T]^T$, $k\in \{1,2,\cdots,n_{tu}\}$.  
Therefore, $rank(\mathbf{D})=rank(\mathbf{D}_k)$.
Any matrix in $\triangle\mathcal{D}$ which does not have 
the above constraint can be obtained by replacing one or more zero 
sub-matrices by non-zero sub-matrices. Since a rank of a matrix will 
not reduce if some of its zero rows/columns are replaced by non-zero 
rows/columns, the minimum rank is obtained by matrices with the above 
constraint.
Let $\triangle\mathbb{A}$ denote the set of non-zero difference QAM/PSK 
constellation points, given by 
$\{s_1-s_2: s_1,s_2 \in \mathbb{A}, s_1 \neq s_2\}$.  
Note that every column of $\mathbf{D}_k$ is either from the set 
$\mathcal{E}_l\Define \{c \hspace{0.5mm}\mathbf{e}_l: c \in \triangle\mathbb{A}\}$ 
for $1\leq l\leq |\mathbb{S}_{\scriptsize\mbox{all}}|$ or from the set 
$\mathcal{E}_{l,q}\Define \{s_1\mathbf{e}_l-s_2\mathbf{e}_q: s_1,s_2 \in \mathbb{A}\}$ 
for $1\leq l\neq q\leq |\mathbb{S}_{\scriptsize\mbox{all}}|$. 
Now, using Proposition 2 of \cite{sm_edas}, the minimum rank of  
$\mathbf{D}_k$ is  
$|\mathbb{S}_{\scriptsize\mbox{all}}|-|\mathbb{S}_{\scriptsize\mbox{sub}}|+1$.
Hence,
$p=rank\left(\mathbf{D}_{\tiny \mbox{min}}\right)\geq \text{min} \{rank\left(\mathbf{D}\right): \mathbf{D} \in \triangle\mathcal{D}\}=|\mathbb{S}_{\scriptsize\mbox{all}}|-|\mathbb{S}_{\scriptsize\mbox{sub}}|+1$. Therefore, $d$ is lower 
bounded by 
$n_rp\geq n_r\left(|\mathbb{S}_{\scriptsize\mbox{all}}|-|\mathbb{S}_{\scriptsize\mbox{sub}}|+1\right)$, i.e., 
\begin{eqnarray}
d &\geq & n_r\left(|\mathbb{S}_{\scriptsize\mbox{all}}|-|\mathbb{S}_{\scriptsize\mbox{sub}}|+1\right).
\label{sys_prop1_lb}
\end{eqnarray}

\subsubsection{Upper bound on $d$}
Consider a pair of transmitted vectors $\mathbf{x}$, $\hat{\mathbf{x}}$
such that 
$\mathbf{x}_1 \neq \hat{\mathbf{x}}_1$, $\mathbf{x}_1=s_1\mathbf{e}_{1}, \hat{\mathbf{x}}_1=\hat{s}_1\mathbf{e}_1$, $\mathbf{x}_i =\hat{\mathbf{x}}_i$, for $2\leq i \leq n_{tu}$, 
where $\mathbf{x}_1, \hat{\mathbf{x}}_1, \mathbf{x}_i$, and 
$\hat{\mathbf{x}}_i$ are defined as in (\ref{gsm_set}). Since we are 
selecting 
$|\mathbb{S}_{\tiny \mbox{sub}}|$ out of $|\mathbb{S}_{\tiny \mbox{all}}|$ 
MAPs, there exist at least one 
$l_{1i}, i=1,\cdots,|\mathbb{S}_{\tiny \mbox{sub}}|$ such that 
$1 \leq l_{1i} \leq p_d$, where $p_d=|\mathbb{S}_{\scriptsize\mbox{all}}|-|\mathbb{S}_{\scriptsize\mbox{sub}}|+1$. Now, we have
\begin{eqnarray}
\| \mathbf{H}\mathbf{A}_{\mathbb{L}_{{\tiny \mbox{ED}}}}\left(\mathbf{x}-\hat{\mathbf{x}}\right)\|^2 \hspace{-3mm}&=&\hspace{-3mm} \| \mathbf{H}^1\mathbf{A}_{\mathbb{L}^1_{{\tiny \mbox{ED}}}}\left(\mathbf{x}_1-\hat{\mathbf{x}}_1\right)\|^2 \label{sys_prop1_6}\\
&\leq &\hspace{-3mm} \max_{\mathbb{L}^1 \in \mathcal{I}^1} \| \mathbf{H}^1\mathbf{A}_{\mathbb{L}^1}\left(\mathbf{x}_1-\hat{\mathbf{x}}_1\right)\|^2 \label{sys_prop1_7}\\
&=& \hspace{-3mm}|s_1-\hat{s}_1|^2 \max_{1 \leq l_{1i} \leq p_d} \|\mathbf{h}_{l_{1i}}^1\|^2 \nonumber\\
& \leq & \hspace{-3mm}|s_1-\hat{s}_1|^2  \sum\limits_{k=1}^{p_d}\sum\limits_{i=1}^{n_r} |h_{i,k}^1|^2. 
\label{sys_prop1_8}
\end{eqnarray}
Note that the set $\mathbb{L}^1_{{\tiny \mbox{ED}}}$ in (\ref{sys_prop1_6}) 
is dependent on $\mathbf{H}$, whereas the set $\mathbb{L}^1$ in 
(\ref{sys_prop1_7}) is independent of $\mathbf{H}$. The unconditional 
PEP between $\mathbf{x}$ and $\hat{\mathbf{x}}$ is given by
\begin{eqnarray}
\hspace{-5mm}P\left(\mathbf x\rightarrow\hat{\mathbf{x}}\right)\hspace{-2.5mm}&= &\hspace{-2.5mm}\mathbb{E}_{\mathbf H}\left\lbrace Q\left(\sqrt{\| \mathbf{H}\mathbf{A}_{\mathbb{L}_{{\tiny \mbox{ED}}}}\left(\mathbf{x}-\hat{\mathbf{x}}\right)\|^2/2\sigma^2}\right)
\right\rbrace \nonumber\\
&\hspace{-30mm}=&\hspace{-17mm} \mathbb{E}_{\mathbf H}\left\lbrace \frac{1}{\pi}\int_{\theta=0}^{\pi/2} \text{exp}\left(-\frac{\| \mathbf{H}\mathbf{A}_{\mathbb{L}_{{\tiny \mbox{ED}}}}\left(\mathbf{x}-\hat{\mathbf{x}}\right)\|^2}{4\sigma^2\sin^2(\theta)}\right) d\theta \right\rbrace\nonumber\\
&\hspace{-30mm}\geq & \hspace{-17mm}\frac{1}{\pi}\int_{\theta=0}^{\pi/2}\hspace{-2mm} \mathbb{E}_{\mathbf H}\Bigg\lbrace\hspace{-0.5mm}\text{exp}\Bigg(\hspace{-1mm}\frac{-|s_1-\hat{s}_1|^2}{4\sigma^2\sin^2(\theta)}\sum\limits_{k=1}^{p_d}\sum\limits_{i=1}^{n_r} |h_{i,k}^1|^2\Bigg)  \Bigg\rbrace d\theta. \label{sys_prop1_9}
\end{eqnarray}
Since $|h_{i,k}^1|^2$s are independent and exponentially distributed with unit mean, we have
\begin{eqnarray}
P\left(\mathbf x\rightarrow\hat{\mathbf{x}}\right)\hspace{-2mm}&\geq&\hspace{-2mm}\frac{1}{\pi}\int_{\theta=0}^{\pi/2} \left(1+\frac{|s_1-\hat{s}_1|^2}{4\sigma^2\sin^2(\theta)}\right) ^{-n_rp_d} d\theta. \label{sys_prop1_10}
\end{eqnarray}
Since $\frac{|s_1-\hat{s}_1|^2}{4\sigma^2\sin^2(\theta)} \gg 1$ at high SNRs, 
we can write
\begin{eqnarray}
P\left(\mathbf x\rightarrow\hat{\mathbf{x}}\right)\geq \left(\frac{|s_1-\hat{s}_1|^2}{4\sigma^2}\right)^{-n_rp_d}\frac{1}{\pi}\int_{\theta=0}^{\pi/2} \sin^{2n_rp_d}(\theta)  d\theta, \label{sys_prop1_11}
\end{eqnarray}
which shows the diversity order of $P\left(\mathbf x\rightarrow\hat{\mathbf{x}}\right)$ is upper  bounded by $n_rp_d=n_r\left(|\mathbb{S}_{\scriptsize\mbox{all}}|-|\mathbb{S}_{\scriptsize\mbox{sub}}|+1\right)$, i.e., $d_{\mathbf{x},\hat{\mathbf{x}}}\leq n_r\left(|\mathbb{S}_{\scriptsize\mbox{all}}|-|\mathbb{S}_{\scriptsize\mbox{sub}}|+1\right)$.  From (\ref{div_d}), an upper bound on $d$ is 
obtained as 
\begin{eqnarray}
d = \min_{\mathbf{x} \neq \tilde{\mathbf{x}}} d_{\mathbf{x},\tilde{\mathbf{x}}} \ \leq \ 
d_{\mathbf{x},\hat{\mathbf{x}}} \ \leq \ n_r\left(|\mathbb{S}_{\scriptsize\mbox{all}}|-|\mathbb{S}_{\scriptsize\mbox{sub}}|+1\right).
\label{sys_prop1_ub}
\end{eqnarray}
Finally, from (\ref{sys_prop1_lb}) and (\ref{sys_prop1_ub}), we see 
that the diversity order $(d)$ achieved by the ED-based MAP selection 
scheme is  
$n_r\left(|\mathbb{S}_{\scriptsize\mbox{all}}|-|\mathbb{S}_{\scriptsize\mbox{sub}}|+1\right)$.

\section*{Appendix B}
\section*{Proof of Proposition 2} 
{\em Proof:} 
\setcounter{subsubsection}{0}
Let $k^*$ be the solution to the optimization problem 
in (\ref{opt_pccr}) for a given realization $\tilde{\mathbf{h}}$, i.e., the 
$k^*$th receive antenna is selected. Let $d_{\scriptsize{\mbox{pc-cr}}}$ 
denote the diversity order achieved by PC-CR scheme.
In the following, we derive upper and lower bounds on 
$d_{\scriptsize{\mbox{pc-cr}}}$ and show that these bounds turn out to
be the same, given by $n_r(n_{tu}+1)$.

\subsubsection{Lower bound on $d_{\scriptsize{\mbox{pc-cr}}}$} 
Let $\triangle {\mathbb V}_{\scriptsize\mbox{pc-cr}}^{{k}}$ be the 
set of difference vectors corresponding to the set 
${\mathbb V}_{\scriptsize\mbox{pc-cr}}^{{k}}$, i.e., 
$\triangle{\mathbb V}_{\scriptsize\mbox{pc-cr}}^{k}=\{\mathbf{v}_1-\mathbf{v}_2: \mathbf{v}_1, \mathbf{v}_2\in{\mathbb V}_{\scriptsize\mbox{pc-cr}}^{{k}}, \mathbf{v}_1 \neq \mathbf{v}_2\}$. 
Let $\triangle\mathcal{D}$ be the set of matrices defined as
{\small
$ \triangle\mathcal{D}\hspace{-0.5mm}= \hspace{-0.5mm}\left\{\mathbf{D}=[\mathbf{B}_1\mathbf{d}_1 \hspace{1.5mm} \mathbf{B}_2\mathbf{d}_2 \hspace{0.5 mm} \cdots \hspace{0.5mm} \mathbf{B}_{n_r}\mathbf{d}_{n_r}]: 
\mathbf{d}_k \in \triangle{\mathbb V}_{\scriptsize\mbox{pc-cr}}^{{k}},  k=1, \cdots, n_r\right\}$. 
}
The size of each matrix in $\triangle\mathcal{D}$ is 
$n_rn_{tu}N_m \times n_r$. The conditional BEP at high SNRs is given by
\begin{eqnarray}
\hspace{-6mm}P_{B|\tilde{\mathbf{h}}} \hspace{-3mm} &\approx & \hspace{-3mm} P_{B|\tilde{\mathbf{h}}_{k^*}} 
\hspace{-1.0mm} \leq \hspace{-0.0mm}Q\hspace{-0.5mm}\Bigg(\hspace{-1.5mm}\bigg(
{\min\limits_{\substack{\mathbf{v}_1, \mathbf{v}_2\in {\mathbb{V}_{\scriptsize\mbox{pc-cr}}^{k^*}}\\{\mathbf v_1}\neq{\mathbf v_2}}}} \hspace{-3mm}\frac{ |\tilde{\mathbf{h}}^T\mathbf{B}_{k^*}\hspace{-0.5mm}\left(\mathbf{v}_1-{\mathbf{v}_2}\right)|^2}{2\sigma^2}\bigg)^{\hspace{-0.5mm}{\frac{1}{2}}}\hspace{-0.5mm}\Bigg)\hspace{-0.0mm}.
\label{eqn29}
\end{eqnarray}
The inequality in (\ref{eqn29}) follows from 
$1 \leq {\delta(\mathbf v_1, \mathbf v_2)} \leq \eta$.  Let 
$\mathbf{d}^{\scriptsize{\mbox{min}}}_k=\argmin_{\mathbf{d}\in \triangle 
{\mathbb V}_{\scriptsize\mbox{pc-cr}}^{{k}}} |\tilde{\mathbf{h}}^T\mathbf{B}_k\mathbf{d}|^2$
for $1\leq k \leq n_r$ denote the difference  vector from $\triangle 
{\mathbb V}_{\scriptsize\mbox{pc-cr}}^{{k}}$ corresponding to the minimum ED.
We then have 
\begin{eqnarray}
\hspace{-5.5mm}P_{B|\tilde{\mathbf{h}}} \leq  Q\bigg(\hspace{-1.5mm}\sqrt{\frac{|\tilde{\mathbf{h}}^T\mathbf{B}_{k^*}{\mathbf{d}^{\scriptsize{\mbox{min}}}_{k^*}}|^2}{2\sigma^2}}\hspace{0mm}\bigg)\hspace{-3.5mm}&\leq &\hspace{-3mm}\frac{1}{2}\text{exp}\bigg(\frac{-|\tilde{\mathbf{h}}^T\mathbf{B}_{k^*}{\mathbf{d}^{\scriptsize{\mbox{min}}}_{k^*}}|^2}{4\sigma^2}\bigg) \label{chernoff} \\
&\hspace{-30mm}\leq &\hspace{-17mm}\frac{1}{2} \text{exp}\Big(\frac{-1}{4n_r\sigma^2}\sum_{i=1}^{n_r}| \tilde{\mathbf{h}}^T\mathbf{B}_{i}{\mathbf{d}^{\scriptsize{\mbox{min}}}_{i}}|^2\Big)\label{max_Ed}\\
&\hspace{-30mm}=&\hspace{-17mm} \frac{1}{2}\prod_{i=1}^{n_r}\text{exp}\Big(-\frac{1}{4\hat{\sigma}^2}| \tilde{\mathbf{h}}_i^T{\mathbf{d}^{\scriptsize{\mbox{min}}}_{i}}|^2\Big),
\nonumber
\end{eqnarray}
where $\hat{\sigma}^2=n_r\sigma^2$. The second inequality in (\ref{chernoff}) 
follows from Chernoff bound and the inequality in (\ref{max_Ed}) is due 
to the fact that ${\mathbf{d}^{\scriptsize{\mbox{min}}}_{k^*}}$ 
corresponds to the maximum ED among the elements  
$\{\mathbf{d}_k^{\scriptsize{\mbox{min}}}\}_{k=1}^{n_r}$.
The average BEP at high SNRs is then given by 
\begin{eqnarray}
P_B \ \approx \ \mathbb{E}_{\tilde{\mathbf{h}}}\{ P_{B|\tilde{\mathbf{h}}}\} 
\hspace{-3mm}&\leq &\hspace{-3mm}  \frac{1}{2}\mathbb{E}_{\tilde{\mathbf{h}}}\bigg\lbrace\prod_{i=1}^{n_r}\text{exp}\bigg(-\frac{1}{4\hat{\sigma}^2}| \tilde{\mathbf{h}}_i^T{\mathbf{d}^{\scriptsize{\mbox{min}}}_{i}}|^2\bigg) \bigg\rbrace\nonumber\\
&\hspace{-45mm}= & \hspace{-25mm}\frac{1}{2}\Bigg(\hspace{-1mm} \underbrace{ \mathbb{E}_{\tilde{\mathbf{h}}_1}\bigg\lbrace\text{exp}\bigg(\frac{-1}{4\hat{\sigma}^2}| \tilde{\mathbf{h}}_1^T{\mathbf{d}^{\scriptsize{\mbox{min}}}_{1}}|^2\bigg)\hspace{-1mm} \bigg\rbrace}_{\triangleq P_{BU}}\hspace{-1mm}\Bigg)^{n_r} 
\hspace{-1.5mm} = \hspace{-0.5mm} \frac{1}{2}\left({P_{BU}}\right)^{n_r}\hspace{-1mm},
\label{iid3}
\end{eqnarray}
where the equality in (\ref{iid3}) follows from the 
independent and identical distribution of  $\tilde{\mathbf{h}}_i$s. 
Next, we show that the diversity order of 
$P_{BU}$ is lower bounded by $n_{tu}+1$.

\subsubsection*{Diversity order of  $P_{BU}$} 
Let $d_{\mathbf{v}_k,\mathbf{v}_{k'}}$ denote the diversity order of PEP 
$P\left({\mathbf v}_k\rightarrow{\mathbf{v}_{k'}}\right)$. The codeword 
error probability (CEP) can be approximated by applying nearest neighbor 
approximation in the high SNR region (\cite{gold_smith}, Eq. 5.45), as 
\begin{eqnarray}
\hspace{-5mm}\hspace{-1mm}\sum_{\mathbf{v}_k}\hspace{-1.5mm}\sum_{\mathbf{v}_{k'}\neq\mathbf{v}_k} \hspace{-3mm}P\left({\mathbf v}_k\rightarrow{\mathbf{v}_{k'}}\right) \hspace{-3mm} &\approx & \hspace{-2.5mm} \mathbb{E}_{\tilde{\mathbf{h}}_1}\hspace{-0mm}\bigg\lbrace \hspace{-0.5mm}Q\hspace{-0.5mm}\bigg(\hspace{-0.5mm}\sqrt{{\frac{|\tilde{\mathbf{h}}_1^T{\mathbf{d}^{\scriptsize{\mbox{min}}}_{1}}|^2}{2\hat{\sigma}^2}}}\bigg)\hspace{-1mm}\bigg\rbrace \hspace{-1mm} \approx \hspace{-1mm} \frac{P_{BU}}{2} .
\label{ber_upper_boun}
\end{eqnarray}
From (\ref{ber_upper_boun}), the diversity order of $P_{BU}$, denoted by 
$d_{BU}$, is given by 
$\min_{{\mathbf v}_k \neq {\mathbf v}_{k'}} d_{\mathbf{v}_k,\mathbf{v}_{k'}}$. 
Let $l_{j_{k}}$ denote the index of the MAP chosen on the $j$th MBM-TU 
in the transmit vector $\mathbf{v}_k$, i.e., $l_{j_{k}}$ is the index of 
the non-zero entry in the vector  
$[v_{(j-1)N_m+1}^k \ \  v_{(j-1)N_m+2}^k \ \ \cdots \ \ v_{jN_m}^k]$. 
Let $\mathbb{G}_{k,k'}^c$ denote the set defined as 
$\mathbb{G}_{k,k'}^c\triangleq \{j:l_{j_{k}}=l_{j_{k'}}, j=1,\cdots,n_{tu}\}$, 
i.e., set of $j$s for which  $l_{j_{k}}$ and 
$l_{j_{k'}}$ are same. Likewise, let $\mathbb{G}_{k,k'}^d$ 
denote the set defined as  
$\mathbb{G}_{k,k'}^d \triangleq \{j:l_{j_{k}}\neq l_{j_{k'}}, j=1,\cdots,n_{tu}\}$, 
i.e., set of $j$s for which  $l_{j_{k}}$ and $l_{j_{k'}}$ are different. 
Note that, for any $k\neq k'$, 
$|\mathbb{G}_{k,k'}^c|+|\mathbb{G}_{k,k'}^d|=n_{tu}$,
and $0 \leq |\mathbb{G}_{k,k'}^c| \leq n_{tu}-1$. 
Now, the PEP between $\mathbf{v}_k$ and $\mathbf{v}_{k'}$ is given by
\begin{eqnarray}
P(\mathbf{v}_k\rightarrow \mathbf{v}_{k'})\hspace{-2mm}&=&\hspace{-2mm} \mathbb{E}_{\tilde{\mathbf{h}}_1}\bigg\lbrace Q\bigg(\sqrt{| \tilde{\mathbf{h}}_1^T(\mathbf{v}_k-\mathbf{v}_{k'})|^2/{2\hat{\sigma}^2}}\hspace{1mm}\bigg) \bigg\rbrace \nonumber\\
&\hspace{-30mm}=&\hspace{-17mm} \mathbb{E}_{\tilde{\mathbf{h}}_1}\hspace{-1mm}\bigg\lbrace \hspace{-1mm} Q\bigg(\hspace{-1mm}\Big(\frac{1}{2\hat{\sigma}^2}{\big|\sum\limits_{j=1}^{n_{tu}}| h_{1,l_{j_k}}^j|e^{\imath{\psi_{k}}}-| h_{1,l_{j_{k'}}}^j|e^{\imath{\psi_{k'}}}\big|^2}\Big)^{\frac{1}{2}}\hspace{-0mm}\bigg)\hspace{-1mm} \bigg\rbrace\nonumber\\
&\hspace{-30mm}=& \hspace{-17mm} \mathbb{E}_{\tilde{\mathbf{h}}_1}\Big\lbrace Q\Big(\sqrt{\left({t_c+t_d+\Delta}\right)/{2\hat{\sigma}^2}}\hspace{1mm}\Big)\Big\rbrace,
\label{eqx1}
\end{eqnarray}
where {\small $t_c\hspace{-1mm}=\hspace{-1mm}S_cP_c, t_d \hspace{-1mm}= \hspace{-1mm}S_d^{k}+S_d^{k'}-2S_k^d S_{k'}^d\cos(\psi_{k,k'}), \Delta=2\big(S^c\big(S_k^d\hspace{-1mm}+\hspace{-1mm}S_{k'}^d) \big(1-\cos(\psi_{k,k'})\big)+\sum\limits_{j \in \mathbb{G}_{k,k'}^c}\sum\limits_{j'\neq j \in \mathbb{G}_{k,k'}^c}\hspace{-2mm} |h_{1,l_{j_k}}^j||h_{1,l_{{j'}_k}}^{j'}|P_c+\sum\limits_{j \in \mathbb{G}_{k,k'}^d}\sum\limits_{j'\neq j \in \mathbb{G}_{k,k'}^d} \hspace{-2mm}\big( |h_{1,l_{j_k}}^j||h_{1,l_{{j'}_k}}^{j'}|+|h_{1,l_{j_{k'}}}^j||h_{1,l_{{j'}_{k'}}}^{j'}|\big)\big), S_c=\sum\limits_{j \in \mathbb{G}_{k,k'}^c}| h_{1,l_{j_k}}^j|^2, P_c=| e^{\imath{\psi_{k}}}-e^{\imath{\psi_{k'}}}|^2, S_d^k=\sum\limits_{j \in \mathbb{G}_{k,k'}^d}  \hspace{-3mm}| h_{1,l_{j_k}}^j|^2, S_d^{k'}= \hspace{-2mm}| h_{1,l_{j_{k'}}}^j|^2, S_k^d\hspace{0mm}=\hspace{-3mm}\sum\limits_{j \in \mathbb{G}_{k,k'}^d}\hspace{-3mm}| h_{1,l_{j_k}}^j|, S_{k'}^d\hspace{-0mm}= \hspace{-3mm}\sum\limits_{j \in \mathbb{G}_{k,k'}^d} \hspace{-3mm} | h_{1,l_{j_{k'}}}^j|, S^c=\sum\limits_{j \in \mathbb{G}_{k,k'}^c}\hspace{-3mm}| h_{1,l_{j_k}}^j|$, and $\psi_{k,k'}=\psi_{k}-\psi_{k'}$.} 
Since $\Delta \hspace{-0mm}\geq \hspace{-0mm}0$, we have
\begin{eqnarray}
\hspace{-0mm}P(\mathbf{v}_k\rightarrow \mathbf{v}_{k'}) &\hspace{-1mm}\leq& \hspace{-1mm} \mathbb{E}_{\tilde{\mathbf{h}}_1}\Big\lbrace Q\Big(\sqrt{\left(t_c+t_d\right)/{2\hat{\sigma}^2}}\hspace{1mm}\Big) \Big\rbrace \label{eqn39}\nonumber\\
&\leq &  \mathbb{E}_{\tilde{\mathbf{h}}_1}\Big\lbrace \frac{1}{2}\text{exp}\Big(-\left(t_c+t_d\right)/{4\hat{\sigma}^2}\Big) \Big\rbrace\label{eqn40}\\
&\hspace{-40mm}=&\hspace{-22mm}  \frac{1}{2}\underbrace{\mathbb{E}_{\tilde{\mathbf{h}}_1}\big\lbrace \text{exp}\big(\frac{-t_c}{4\hat{\sigma}^2}\big)\big\rbrace}_{\triangleq \ T_c}\underbrace{\mathbb{E}_{\tilde{\mathbf{h}}_1}\big\lbrace \text{exp}\big(\frac{-t_d}{4\hat{\sigma}^2}\big)\big\rbrace}_{\triangleq \ T_d}=\frac{1}{2}T_c T_d,
\label{eqn41}
\end{eqnarray}
where the inequality in (\ref{eqn40}) follows from Chernoff bound and the 
equality in (\ref{eqn41}) follows from independence of  
$\{h_{1,l_{j_k}}^j \hspace{-2mm} \in \hspace{-0mm} \mathbb{G}_{k,k'}^c\}$ and $\{h_{1,l_{j_k}}^j,h_{1,l_{j_{k'}}}^j \hspace{-2mm} \in \hspace{-0mm} \mathbb{G}_{k,k'}^d$\}.   
Now, since $|h_{1,l_{j_k}}^j|^2$s are independent and exponentially
distributed with unit mean, from  (\ref{eqn41}), 
$T_c$ can be written as 
\begin{eqnarray*}
T_c  = \mathbb{E}_{\tilde{\mathbf{h}}_1}\bigg \lbrace \text{exp}\bigg(\frac{-P_c}{4\hat{\sigma}^2}\hspace{-2mm}\sum\limits_{j \in \mathbb{G}_{k,k'}^c}\hspace{-2.5mm}| h_{1,l_{j_k}}^j|^2\bigg)\bigg\rbrace \hspace{-1mm}=\hspace{-1mm}
\left(\hspace{-0mm}1+\frac{P_c}{4\hat{\sigma}^2}\hspace{-0mm}\right)^{-|\mathbb{G}_{k,k'}^c|}.
\end{eqnarray*}
Since $\frac{P_c}{4\hat{\sigma}^2} \gg 1$ at high SNRs, we can approximate 
$T_c$ as
\begin{eqnarray}
T_c \hspace{-1mm}&\approx &\hspace{-1mm} {(P_c/4)}^{-|\mathbb{G}_{k,k'}^c|}\big(1/{\hat{\sigma}^2}\big)^{-(|\mathbb{G}_{k,k'}^c|)}.
\label{pep_tc}
\end{eqnarray}
Similarly,  from (\ref{eqn41}), $T_d$ can be written as
\begin{eqnarray}
\hspace{-25mm} T_d\hspace{-2.5mm}&= &\hspace{-2.5mm} \mathbb{E}_{\tilde{\mathbf{h}}_1}\hspace{-1mm}\Big\lbrace \hspace{-0mm}\text{exp}\hspace{-0mm}\Big(\hspace{-0mm}\frac{-1}{4\hat{\sigma}^2}\hspace{-0mm}\big(\hspace{-0mm}S_d^k+S_d^{k'}-2S_k^dS_{k'}^d\cos(\psi_{k,k'})\hspace{-0mm}\big) \hspace{-0.5mm}\Big)\hspace{-0.5mm} \Big\rbrace \nonumber \\ 
&\hspace{-13.5mm}\leq&\hspace{-10mm} \mathbb{E}_{\tilde{\mathbf{h}}_1}\hspace{-1mm}\Big\lbrace \hspace{-0.5mm}\text{exp}\hspace{-0mm}\Big(\hspace{-0mm} \frac{-1}{4\hat{\sigma}^2} \big(\hspace{-0.5mm} S_d^k\hspace{-0.5mm}+\hspace{-0.5mm}S_d^{k'}\hspace{-2mm}-\hspace{-1mm}2|\mathbb{G}_{k,k'}^d|\big(\hspace{-0.5mm}S_d^kS_d^{k'}\hspace{-0.5mm}\big)^{\frac{1}{2}}\hspace{-1mm}\cos(\hspace{-0.5mm}\psi_{k,k'}\hspace{-0.5mm})\big) \hspace{-1.2mm}\Big)\hspace{-1.2mm} \Big\rbrace,
\label{eqn_prop2_1}
\end{eqnarray}
where the inequality in (\ref{eqn_prop2_1}) follows from Cauchy-Schwartz 
inequalities given by 
{\small $S_k^d\leq \big(| \mathbb{G}_{k,k'}^d|S_d^k\big)^{\frac{1}{2}}$} 
and 
{\small $S_{k'}^d\leq \big(| \mathbb{G}_{k,k'}^d|S_d^{k'}\big)^{\frac{1}{2}}$}.
Let $p=(S_d^k)^{0.5}$ and $q=(S_d^{k'})^{0.5}$. 
Note that $p$ and $q$  are central Chi distributed with 
$2| \mathbb{G}_{k,k'}^d|$ degrees of freedom. Therefore, (\ref{eqn_prop2_1}) 
can be written  as
\begin{eqnarray*}
\hspace{0.5mm}T_d \hspace{-1mm}\leq \hspace{-2mm}\int_{p=0}^{\infty}\hspace{-1mm}\int_{q=0}^{\infty}\hspace{-3mm}e^{\frac{-{(p^2+q^2\hspace{-0.5mm}-\hspace{-0.5mm}2pq\cos(\psi_{k,k'}\hspace{-0.5mm})\hspace{-0.5mm})}}{4\hat{\sigma}^2}}\hspace{-1mm}\frac{(pq)^{2| \mathbb{G}_{k,k'}^d|-1}\hspace{-0.5mm}e^{\frac{-(p^2+q^2)}{2}}}{4^{|\mathbb{G}_{k,k'}^d|-1}\hspace{-0.5mm}(\Gamma(| \mathbb{G}_{k,k'}^d|))^2}dp \hspace{0.5mm}dq,
\end{eqnarray*}
where $\Gamma\left(.\right)$ denotes the Gamma function. Substituting 
$p=r\cos\theta$ and $q=r\sin\theta$, we have 
\begin{eqnarray}
T_d \hspace{-3mm}&\leq & \hspace{-4mm}\int_{\theta=0}^{\frac{\pi}{2}}\hspace{-1mm}\frac{8\sin ^{2| \mathbb{G}_{k,k'}^d|-1}(2\theta)}{\Gamma(| \mathbb{G}_{k,k'}^d|)^22^{4| \mathbb{G}_{k,k'}^d|}}\hspace{-0.5mm}\int_{r=0}^{\infty}\hspace{-4mm}{\substack{ e^{-\frac{r^2}{4\hat{\sigma}^2}\big(1+2\hat{\sigma}^2-\sin(2\theta)\cos(\psi_{k,k'})\big)} \\ r^{4\big| \mathbb{G}_{k,k'}^d\big|-1}dr \hspace{0.5mm}d{\theta}}}  \nonumber\\
&\hspace{-8mm}=&\hspace{-8.5mm}\int_{\theta=0}^{\frac{\pi}{2}}\hspace{-3mm}4\frac{\Gamma({2| \mathbb{G}_{k,k'}^d|})}{\Gamma(| \mathbb{G}_{k,k'}^d|)^2}\hspace{-0.5mm}{\left(\hspace{-0.5mm}\frac{1\hspace{-1mm}+\hspace{-1mm}2\hat{\sigma}^2\hspace{-1mm}-\hspace{-1mm}\sin(2\theta)\hspace{-0.5mm}\cos(\psi_{k,k'})}{\sin(2\theta)\hat{\sigma}^2}\hspace{-1.2mm}\right)^{\hspace{-1.5mm}-2| \mathbb{G}_{k,k'}^d|}}\hspace{-10mm}\sin ^{-1}(2\theta)\hspace{0.5mm}d{\theta}.\nonumber
\end{eqnarray}
Since $1 \gg {2\hat{\sigma}^2}$ at high SNRs, we can write 
\begin{eqnarray}
\hspace{-10mm}T_d \hspace{-2mm}&\leq &\hspace{-2mm} \big(1/{\hat{\sigma}^2}\big)^{-2| \mathbb{G}_{k,k'}^d|} \nonumber\\
& & \hspace{-15mm}\underbrace{\int_{\theta=0}^{\frac{\pi}{2}}\hspace{-2.5mm}4
\frac{\Gamma({2| \mathbb{G}_{k,k'}^d|})}{\Gamma(| \mathbb{G}_{k,k'}^d|)^2}
\hspace{-0mm}{\left(\hspace{-1mm}\frac{1\hspace{-1mm}-\hspace{-1mm}\sin(2\theta)\hspace{-0.5mm}\cos(\psi_{k,k'})}{\sin(2\theta)}\hspace{-1mm}\right)^{\hspace{-1.5mm}-2| \mathbb{G}_{k,k'}^d|}}\hspace{-7mm}\sin ^{-1}(2\theta)\hspace{0.5mm}d{\theta}
}_{\triangleq k_d}. 
\label{pep_td}
\end{eqnarray}
Note that $k_d$ is independent of $\hat{\sigma}^2$. 
Now, from (\ref{eqn41}), (\ref{pep_tc}), (\ref{pep_td}), we can write 
\begin{eqnarray}
\hspace{-5mm}P(\mathbf{v}_k\rightarrow \mathbf{v}_{k'}) &\hspace{-3mm}\leq& \hspace{-3mm} \frac{k_d}{2}\Big(\frac{P_c}{4}\Big)^{-|\mathbb{G}_{k,k'}^c|}\Big(\frac{1}{\hat{\sigma}^2}\Big)^{-(|\mathbb{G}_{k,k'}^c|+2|\mathbb{G}_{k,k'}^d|)},
\label{eqn42}
\end{eqnarray}
which shows that the diversity order  
$d_{\mathbf{v}_k,  \mathbf{v}_{k'}} \geq |\mathbb{G}_{k,k'}^c|+2|\mathbb{G}_{k,k'}^d|$. 
Hence, the diversity order of $P_{BU}$ is 
$d_{BU}=\min_{\mathbf{v}_{k}\neq\mathbf{v}_{k'}} d_{\mathbf{v}_k,  \mathbf{v}_{k'}} \geq  \min_{\mathbf{v}_{k}\neq\mathbf{v}_{k'}}|\mathbb{G}_{k,k'}^c|+2|\mathbb{G}_{k,k'}^d|$. Since  $|\mathbb{G}_{k,k'}^c|+|\mathbb{G}_{k,k'}^d|=n_{tu}$
for any $k\neq k'$, the minimum value for 
$|\mathbb{G}_{k,k'}^c|+2|\mathbb{G}_{k,k'}^d|=2n_{tu}-|\mathbb{G}_{k,k'}^c|$
is obtained when $|\mathbb{G}_{k,k'}^c|$ is maximum. That is,	
$\min_{k\neq k'} |\mathbb{G}_{k,k'}^c|+2|\mathbb{G}_{k,k'}^d| = 2n_{tu} - \max_{k\neq k'} |\mathbb{G}_{k,k'}^c|$.
Also, since $0 \leq |\mathbb{G}_{k,k'}^c| \leq n_{tu}-1$ 
for any $k\neq k'$,
$\max_{k\neq k'} |\mathbb{G}_{k,k'}^c| = n_{tu}-1$. Hence,
$\min_{k\neq k'} |\mathbb{G}_{k,k'}^c|+2|\mathbb{G}_{k,k'}^d| = 
2n_{tu} - (n_{tu}-1) = n_{tu}+1$, which gives $d_{BU}\geq n_{tu}+1$.
Therefore, from (\ref{iid3}), the diversity order achieved by 
the PC-CR scheme is lower bounded as 
\begin{eqnarray} 
d_{\scriptsize{\mbox{pc-cr}}} \ \geq \ n_rd_{BU} \ \geq \ n_r(n_{tu}+1). 
\label{sys_prop2_lb}
\end{eqnarray}

\subsubsection{Upper bound on $d_{\scriptsize{\mbox{pc-cr}}}$}
Consider a pair of transmitted vectors $\mathbf{v}_k, \hat{\mathbf{v}}_{k'}$ 
such that $|\mathbb{G}_{k,k'}^c| = n_{tu}-1$. Without loss of generality, 
assume $\mathbb{G}_{k,k'}^c =\{1,\ 2, \ \cdots, \ n_{tu}-1\}$. We have
\begin{eqnarray}
\hspace{-7mm}
| \tilde{\mathbf{h}}_{k^*}^T(\mathbf{v}_k-\hat{\mathbf{v}}_{k'})|^2& \hspace{-2mm} \leq & \hspace{-2mm} \sum\limits_{i=1}^{n_r}| \tilde{\mathbf{h}}_i^T(\mathbf{u}_ke^{\imath{\psi_{k}}}-\hat{\mathbf{u}}_{k'}e^{\imath{\psi_{k'}}})|^2 \nonumber\\
&\hspace{-48mm}\leq & \hspace{-25mm}\sum\limits_{i=1}^{n_r}| \tilde{\mathbf{h}}_i^T(\mathbf{u}_k+\hat{\mathbf{u}}_{k'})|^2 \label{sys_prop2_7}\\
&\hspace{-48mm}=&\hspace{-25mm}\sum\limits_{i=1}^{n_r}\big( 2\sum\limits_{j=1}^{n_{tu}-1}| h_{i,l_{j_k}}^j|+| h_{1,i_{{n_{tu}}_k}}^{n_{tu}}|+| h_{i,l_{{n_{tu}}_{k'}}}^{n_{tu}}|\big)^2 \nonumber\\
&\hspace{-48mm}\leq &\hspace{-25mm} (4n_{tu}-2)\hspace{-1mm}\sum\limits_{i=1}^{n_r} \hspace{-1mm}\sum\limits_{j=1}^{n_{tu}-1}\hspace{-2mm}| h_{i,l_{j_k}}^j|^2\hspace{-1mm}+\hspace{-1mm}| h_{i,l_{{n_{tu}}_k}}^{n_{tu}}|^2\hspace{-1mm}+\hspace{-1mm}| h_{i,l_{{n_{tu}}_{k'}}}^{n_{tu}}|^2,
\label{sys_prop2_8}
\end{eqnarray}
where the inequality in (\ref{sys_prop2_7}) follows from the fact that the 
distance between two points is maximum when phase difference between them 
is $\pi$, and the inequality in (\ref{sys_prop2_8}) follows from 
Cauchy-Schwartz inequality. Now, using (\ref{sys_prop2_8}) and following 
similar derivation steps in Eqs. (\ref{sys_prop1_9})-({\ref{sys_prop1_11}) 
in Appendix A, the unconditional PEP between $\mathbf{v}_k$ and 
$\hat{\mathbf{v}}_{k'}$,
$P\left(\mathbf v_k \rightarrow{\hat{\mathbf{v}}}_{k'}\right)$, 
can be obtained as
\begin{eqnarray}
P\left(\mathbf v_k \rightarrow{\hat{\mathbf{v}}}_{k'}\right)& = &
\mathbb{E}_{\mathbf H}\left\lbrace Q\left(\sqrt{| \tilde{\mathbf{h}}_{k^*}^T(\mathbf{v}_k-\hat{\mathbf{v}}_{k'})|^2/2\sigma^2}\right)
\right\rbrace \nonumber\\
&\hspace{-40mm} \geq &\hspace{-23mm} \bigg(\frac{4n_{tu}-2}{4\sigma^2}\bigg)^{-n_r\left(n_{tu}+1\right)}\frac{1}{\pi}\int_{\theta=0}^{\frac{\pi}{2}} \sin^{2{n_r\left(n_{tu}+1\right)}}(\theta)  
d\theta, \label{sys_prop1_11a}
\end{eqnarray}
which shows that the diversity order of 
$P\left(\mathbf v_k\rightarrow{\hat{\mathbf{v}}_{k'}}\right)$ is upper  
bounded by $n_r\left(n_{tu}+1\right)$, i.e., 
$d_{\mathbf{v}_k,{\hat{\mathbf{v}}_{k'}}}\leq n_r\left(n_{tu}+1\right)$.   
Therefore, we have
\begin{eqnarray}
d_{\scriptsize{\mbox{pc-cr}}} = \min_{\mathbf{v}_k \neq {{\mathbf{v}}_{k'}}} d_{\mathbf{v}_k,{\mathbf{v}}_{k'}} \leq  d_{\mathbf{v}_k,{\hat{\mathbf{v}}}_{k'}}  \leq  n_r\left(n_{tu}+1\right).
\label{sys_prop2_ub}
\end{eqnarray}
From (\ref{sys_prop2_lb}) and (\ref{sys_prop2_ub}), we see that the diversity
order $(d_{\scriptsize{\mbox{pc-cr}}})$ achieved by the PC-CR scheme is  $ n_r\left(n_{tu}+1\right)$.

\section*{Acknowledgment}
The authors would like to thank Prof. Arogyaswami Paulraj, Stanford
University for the insightful discussions on the use of parasitic elements
in smart antenna systems.

\end{document}